\documentclass[aps,prd,twocolumn,superscriptaddress,showpacs,showkeys]{revtex4}

\usepackage[dvips]{graphicx}
\usepackage{ulem}
\usepackage{amsmath}
\usepackage{float}
\def\slash#1{\not\!#1}

\begin{document}


\title{Parity projection of QCD sum rules for the nucleon}


\author{Keisuke Ohtani}
\email{ohtani.k@th.phys.titech.ac.jp}
\affiliation{Department of Physics, H-27, Tokyo Institute of Technology, Meguro, Tokyo 152-8551, Japan}

\author{Philipp Gubler}
\email{pgubler@riken.jp}
\affiliation{RIKEN Nishina Center, Hirosawa 2-1, Wako, Saitama, 351-0198, Japan}

\author{Makoto Oka}
\email{oka@th.phys.titech.ac.jp}
\affiliation{Department of Physics, H-27, Tokyo Institute of Technology, Meguro, Tokyo 152-8551, Japan}
\affiliation{J-PARC Branch, KEK Theory Center, Institute of Particle and Nuclear Studies, High Energy Accelerator Research
Organization (KEK), 203-1, Shirakata, Tokai, Ibaraki, 319-1106, Japan}

\date{\today}
\begin{abstract}
The nucleon and its negative-parity excited states are examined in a maximum entropy method analysis of QCD sum rules. 
First, we rederive the parity projected sum rules for baryons using ``old-fashioned" correlation functions. 
Doing this, the method is generalized so that higher order operator product expansion (OPE) terms can be calculated unambiguously. 
We then apply this approach to the nucleon channel taking into account all known first order $\alpha_{s}$ 
corrections to the Wilson coefficients of the OPE. 
As these corrections have turned out to be large, we suppress them by using a phase-rotated 
Gaussian kernel. Simultaneously, this phase rotation strongly suppresses the continuum contribution and improves the convergence of the OPE. 
The resulting sum rule has the interesting feature that it is dominated by the term containing the 
chiral condensate of dimension 3. Analyzing this sum rule by the maximum entropy method, we are able to 
extract information of both the positive and negative parity states. 
\end{abstract}


\maketitle

\section{Introduction \label{sec:introduction}}

The properties of baryons have been investigated by QCD sum rule studies already for more than three decades. 
Historically, this method was developed by Shifman \textit{et al.} \cite{Shifman1,Shifman2} and 
applied to baryonic channels by Ioffe \cite{Ioffe}.
Since then, the nucleon sum rules were continuously improved by including 
higher orders in the perturbative Wilson coefficients \cite{Krasnikov,Chung2,Jamin,Ovchinnikov,Shiomi,Sadovnikova} 
or non-perturbative power corrections \cite{Belyaev,Chung2,Leinweber1}.
Furthermore, QCD sum rules also have been used to investigate the nucleon properties 
in nuclear matter \cite{Hatsuda,Cohen,Drukarev} or at finite temperature \cite{Adami}. 
However, even in the vacuum, it has not been an easy task to appropriately extract information on the
nucleonic states because of several still not completely resolved issues. 

Specifically, in analyzing the nucleon sum rules, we face four major problems.
The first problem is that the nucleon operator couples to both positive and negative parity states. 
The contributions of these states can potentially disturb the analysis and 
especially lower the reliability of the extraction of the negative parity states.
To overcome this difficulty, 
Jido \textit{et al.} \cite{Jido} and Kondo \textit{et al.} \cite{Kondo} have constructed certain 
parity projected sum rules which separate the contributions of the positive and negative parity states. 
However, in these analyses, the first order $\alpha_{s}$ corrections and the contributions of the non-perturbative power 
corrections above dimension 6 were not considered. 
Furthermore, some technical problems of the approach proposed in \cite{Jido} were pointed out in \cite{Kondo}, which are 
related to a not fully justified use of the old-fashioned correlator instead of the time-ordered one and ensuing ambiguities in the 
treatment of the poles at zero energy appearing in the higher order terms of the operator product expansion (OPE). 

The second problem is that the
$\alpha_{s}$ corrections tend to be large \cite{Leinweber2}, which seriously put the convergence of the Wilson coefficients as an expansion in $\alpha_{s}$ into question. 
This issue is especially serious for the parity projected sum rules, for which one cannot avoid the perturbative dimension 0 term, 
where the $\alpha_{s}$ corrections are large for all available local interpolating fields. 

The third problem is the large contribution of the continuum, which makes it difficult to investigate the properties of the nucleon and its excited states. 
Indeed, in a previous study \cite{Leinweber2}, it was shown that the Borel window of the chiral even sum rule is only very narrow if it opens at all. 
Although this difficulty can be avoided 
 by only using the chiral odd terms of the OPE, 
the chiral even terms have to be used to separate the positive and negative parity contributions. 

The fourth problem is mainly concerned with the extraction of the lowest negative parity state. Namely, it is not known which interpolating field 
predominantly couples to such a resonance and which one merely to continuum states.  
As we have shown in a previous study \cite{KOhtani}, 
the spectral functions corresponding to different interpolating field
operators can show a distinct behavior. 
Some of them do not contain any pole structures. 

To remedy these difficulties, we construct the parity projected sum rule including 
first order $\alpha_{s}$ corrections directly from the old-fashioned correlation function 
to resolve the problems pointed out in \cite{Kondo}, 
and analyze the nucleon channel by using the maximum entropy method (MEM). 
The MEM analysis of QCD sum rules is a novel approach which does not require the ``pole+ continuum" ansatz 
and has been applied to the $\rho$ meson \cite{Gubler1} and the nucleon channel \cite{KOhtani} in the 
vacuum and to the 
charmonium \cite{Gubler2} and bottomonium \cite{KSuzuki} channels at zero and finite temperature.
However, when using conventional kernels (Borel or Gaussian), 
because of the large $\alpha_{s}$ corrections, 
the perturbative expansion is not under control and it is quite possible 
that higher order corrections spoil the results.
Hence, we construct a new sum rule 
by using a phase-rotated Gaussian kernel, in 
which the $\alpha_{s}$ corrections are suppressed. 
Additionally, this kernel also suppresses the large continuum contribution and thus improves the extraction of 
the nucleon properties through the spectral function. 
Moreover, this sum rule has the interesting feature 
that it is dominated by the chiral condensate term
and thus is suitable for investigating the relation between the splitting of the positive and negative parity states 
and chiral symmetry. 

The paper is organized as follows. 
In Sec. \ref{sec:2}, we introduce the parity projected sum rules, 
discuss the problem of using the old-fashioned correlator for constructing the sum rules and 
explain how this issue can be resolved. 
Then, we apply our method to the nucleon sum rule taking into account all known 
first order $\alpha_{s}$ corrections. 
The maximum entropy method is briefly introduced in Sec. \ref{sec:3}, after which in 
Sec. \ref{sec:4} the results of the analyses are outlined. Summary and 
conclusions are given in Sec. \ref{sec:5}. 

\section{Parity projected baryonic sum rules}
\label{sec:2} 
\subsection{Parity projection of the correlation function}
In QCD sum rules, one usually studies the properties of the time ordered correlation function: 
\begin{equation}
\begin{split}
\Pi(q) = i\int e^{iqx}\langle 0|T[\eta (x)\overline{\eta}(0)]|0\rangle d^{4}x.
\label{eq:correlation function0}
\end{split}
\end{equation}
Here, $\eta$ is an interpolating field constructed from quark and gluon operators to have the same quantum 
numbers as the hadron of interest.
Making use of the analyticity of this correlation function, we can obtain the dispersion relation given below: 
\begin{equation}
\begin{split}
\Pi(q)&=\frac{1}{\pi}\int^\infty _{0} \frac{\mathrm{Im}\Pi (t)}{t-q^2}dt
=\int^\infty _{0} \frac{\rho (t)}{t-q^2}dt,
\label{eq:disp}
\end{split}
\end{equation}
where $\rho(t)$ denotes the hadronic spectral function, which contains contributions from all possible physical states coupling to $\eta$. 
Note that we ignore here possible subtraction terms, which play no role in the following discussion. 
Eq.(\ref{eq:disp}) is the usual starting point for QCD sum rule analyses. 

In baryonic channels, the interpolating field $\eta$ carrying positive parity couples to both 
positive and negative parity baryon states,
 $|n^{+}(q) \rangle$ and $|n^{-}(q) \rangle$:
\begin{equation}
\begin{split}
\langle 0 | \eta(x) | n^+(q) \rangle &= \lambda^n_+ u_+(p) e^{-iqx}, \\ 
\langle 0 | \eta(x) | n^-(q) \rangle &= \lambda^n_- \gamma_5  u_-(p) e^{-iqx}.
\end{split}
\end{equation}
Here, $u_{\pm}(q)$ are Dirac spinors of positive and negative parity and $\lambda^n_{\pm}$ 
correspond to the strength of the coupling of $\eta$ to the state $|n\rangle$.
Hence, the correlator of $\eta$ will 
have the following form: 
\begin{equation}
\begin{split}
\Pi(q) &= i \displaystyle \int d^4x e^{iqx} \langle 0 | T[\eta(x) \overline{\eta}(0)] | 0\rangle \\
&=\displaystyle -\int _{0}^{\infty} \biggl[ \rho_{+}(m) \frac{\slash{q} + m}{q^2 -m^2 +i\epsilon} \\
&\quad \quad \quad \quad  \quad \quad + \rho_{-}(m) \frac{\slash{q} - m}{q^2 -m^2 +i\epsilon}\biggr] dm  \\
&\equiv \slash{q}\Pi_1(q^2) + \Pi_2(q^2).
\end{split}
\label{eq:torder}
\end{equation}
Here, $\rho_{+(-)}(m)$ contains the contributions of only positive (negative) parity states.
When only the sum rule constructed from either $\Pi_1(q^2)$ or $\Pi_{2}(q^{2})$ is used, it is not possible to determine the parity of 
any structure appearing in the respective spectral function without 
additional assumptions, 
as both positive and negative parity states contribute to it.
Thus, the problem of parity projection boils down to consistently disentangling the contributions of 
positive and negative parity to $\Pi_1(q^2)$ and $\Pi_{2}(q^{2})$. 

To separate each parity contribution, 
the ``old-fashioned" correlator in the rest frame ($\vec{q} = 0$) 
was defined in \cite{Jido}, 
\begin{equation}
\begin{split}
\Pi^{\mathrm{old}}(q_0) &= i \int d^{4}x e^{iqx} \theta(x_0) \langle0|T[\eta(x)\overline{\eta}(0)]|0\rangle\Big|_{\vec{q}=0} \\
&\equiv \gamma_0 \Pi_1^{\mathrm{old}}(q_0) + \Pi_2^{\mathrm{old}}(q_0),
\end{split}
\label{eq:oldfash}
\end{equation}
where the essential difference to Eq.(\ref{eq:torder}) is the insertion of the Heaviside step-function $\theta(x_0)$ 
before carrying out the Fourier transform. 
This correlator contains contributions only from states which propagate forward in time. 

Using Eq.(\ref{eq:torder}) and the Heaviside step-function, 
it can be shown that the 
functions $\Pi_1^{\mathrm{old}}(q_0)$ and $\Pi_2^{\mathrm{old}}(q_0)$ contain only poles in the positive $q_0$ 
region and are analytic for $\mathrm{Im}\,q_0 \geq  0$. 
The details of how this can be done are explained in Appendix A. 
Furthermore, by applying the projection operator $\frac{1}{2}(\gamma_0 \pm 1) \equiv P^{\pm}$ to the old-fashioned correlator of Eq.(\ref{eq:oldfash}) 
and taking the trace over the spinor indices, we are able to construct functions that contain only positive or negative 
parity states, as 
\begin{equation}
\begin{split}
\frac{1}{2} \mathrm{Tr} \big[P^{\pm} \Pi^{\mathrm{old}}(q_0)\big] 
&= \Pi_1^{\mathrm{old}}(q_0) \pm \Pi_2^{\mathrm{old}}(q_0) \equiv \Pi^{\pm}(q_0) \\
&= - \displaystyle \int _{0}^{\infty} \rho_{\pm}(m) \frac{1}{q_0 - m + i\epsilon} dm.
\end{split}
\label{eq:ppfinal}
\end{equation}
The imaginary parts of $\Pi^{\pm}(q_0)$ defined above then give the desired parity projected spectral functions:  
\begin{equation}
\frac{1}{\pi}\mathrm{Im}\,\Pi^{\pm} (q_0) = \displaystyle \rho_{\pm}(q_{0}).
\label{eq:Im}
\end{equation}
These can, however, not be calculated directly because perturbation theory is not reliable in the low $q_0$ region, but 
only at $|q_0| \to \infty$.

\subsection{Construction of the sum rules}
Next, we will discuss the construction of the parity projected sum rules. 
For this purpose, we define two expressions describing the correlation function $\Pi^{\pm}(q_0)$. 
The first one is obtained from the OPE at high energy and is thus called $\Pi^{\pm}_{\mathrm{OPE}}(q_0)$ in what follows. 
The second one, denoted as $\Pi^{\pm}_{\mathrm{Phys.}}(q_0)$, is expressed using the hadronic degrees of freedom in the 
physical energy region of $q_0$. 

\begin{figure}[!tbp]
 \begin{center}
  \includegraphics[scale=0.6]{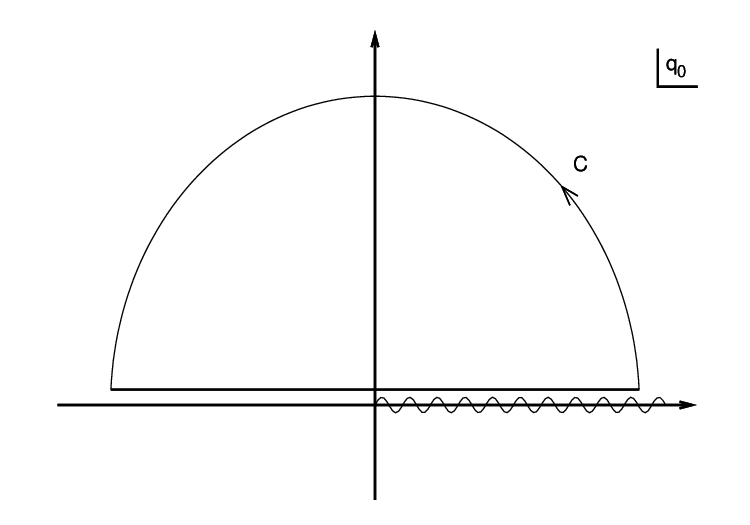}
 \end{center}
 \caption{The contour integral $C$ on the complex plane of the variable $q_0$, used 
in Eq.(\ref{eq:sumrulestart}). For the actual calculations, the radius of the half circle of 
$C$ is taken to infinity. 
The wavy line denotes the non-analytic cut (or poles) of $\Pi^{\pm}(q_0)$ on the 
positive side of the real axis. 
Compared to the discussion in the main text, we have here slightly shifted the contour 
away from the real axis into the upper half of the imaginary plane for better visuality.}
 \label{fig:contour}
\end{figure}

To construct the sum rules, we consider the contour integral 
\begin{equation}
\displaystyle \oint_{\mathrm{C}} dq_0 \big[ \Pi^{\pm}_{\mathrm{OPE}}(q_0) - \Pi^{\pm}_{\mathrm{Phys.}}(q_0) \big] W(q_0) = 0,
\label{eq:sumrulestart}
\end{equation}
where the contour C is given in Fig. \ref{fig:contour}.
$W(q_0)$ must be analytic in the upper half of the imaginary plane and real on the real axis. As long as it satisfies these 
conditions, it can be chosen arbitrarily. That Eq.(\ref{eq:sumrulestart}) gives 0 follows from the analyticity of both 
$\Pi^{\pm}_{\mathrm{OPE}}(q_0)$ and $\Pi^{\pm}_{\mathrm{Phys.}}(q_0)$. In other words, there are no poles or cuts inside 
of the contour $C$. Now, we know from asymptotic freedom that the perturbative expression $\Pi^{\pm}_{\mathrm{OPE}}(q_0)$ 
approaches the hadronic one at $|q_0| \to \infty$. Thus, the integrand of Eq.(\ref{eq:sumrulestart}) vanishes on the half 
circle of $C$, whose radius is taken to infinity. 
Therefore, we are left with the section of $C$ which runs along the real axis and can write down the 
sum rule as 
\begin{equation}
\displaystyle \int_{-\infty}^{\infty} dq_0 \Pi^{\pm}_{\mathrm{OPE}}(q_0) W(q_0) =
\displaystyle \int_{-\infty}^{\infty} dq_0 \Pi^{\pm}_{\mathrm{Phys.}}(q_0) W(q_0).
\label{eq:finalpar1}
\end{equation}
Finally, taking the imaginary part of the above equation we get 
\begin{equation}
\begin{split}
&\displaystyle \int_{-\infty}^{\infty} dq_0 \frac{1}{\pi} \mathrm{Im} \bigl[\Pi^{\pm}_{\mathrm{OPE}}(q_0) \bigr] W(q_0) \\
& \quad \quad \quad = \displaystyle \int_{0}^{\infty} dq_0 \rho^{\pm}_{\mathrm{Phys.}}(q_0) W(q_0),
\end{split}
\label{eq:finalpar}
\end{equation}
where we have made use of the definition 
$\rho^{\pm}_{\mathrm{Phys.}}(q_0) \equiv \frac{1}{\pi} \mathrm{Im} \Pi^{\pm}_{\mathrm{Phys.}}(q_0)$. 
Moreover, on the 
right-hand side, we have exploited the fact that $\Pi^{\pm}_{\mathrm{Phys.}}(q_0)$ only has poles on the positive side 
of the real axis (see Eq.(\ref{eq:ppfinal})) and have hence restricted the integral to this region.

Here, we comment on some important differences between the sum rules constructed 
in this paper and the one considered in the previous study of \cite{Jido}. 
In contrast to Eq.(\ref{eq:finalpar}), the authors of \cite{Jido} have also restricted the integral on its right-hand side to positive values. 
Even though this procedure is roughly correct, it leads to ambiguities in the higher order OPE terms, which 
in the chiral limit have poles at $q_0 = 0$. 
Furthermore, in \cite{Jido} the imaginary part of the time ordered 
correlator (without $\theta(x_0)$) was used instead of $\mathrm{Im} \Pi^{\pm}_{\mathrm{OPE}}(q_0)$, which in principle 
should be derived from the old-fashioned correlator of Eq.(\ref{eq:oldfash}). As was pointed out in \cite{Kondo}, it is not entirely clear whether this prescription 
is justifiable. Therefore, in this study, we implement two essential improvements as compared to \cite{Jido}: 
1) We derive all results directly from the old-fashioned correlator of Eq.(\ref{eq:oldfash}). 
2) We do not restrict the region of integration of the OPE side of Eq.(\ref{eq:finalpar}) 
to positive values and therefore remove the ambiguities that might occur for higher order OPE terms. 

As a last point, we briefly discuss another parity projected sum rule which was proposed in \cite{Kondo}. 
There, the sum rule is constructed from the retarded correlation function, whose spectral function 
 has poles corresponding to positive (negative ) parity states in the positive (negative ) $q_{0}$ region, respectively.
Hence, the contributions of the positive and negative parity states are separated into the different energy regions 
of the integral in Eq.(\ref{eq:finalpar1}). One then can disentangle positive and negative parity states by using 
suitable combinations of several different kernels. In principle, it is also possible to perform an MEM analysis for such 
combined sum rules 
and we have indeed carried out a trial study to do this. 
However, in the actual analysis, it has turned out to be difficult to extract 
information of positive (negative) parity states which is not affected by negative (positive) parity states. 
The reason is that the MEM results have the tendency to be symmetric with respect to 
the positive and negative energy contributions, 
which can cause artificial peaks \cite{Gull}. 
Therefore, to avoid this problem we use the method of \cite{Jido} in this paper. 

\subsection{The parity projected nucleon sum rule}
\label{subsec:2.2} 
In this subsection, we construct the parity projected sum rule of the nucleon including all known 
$\alpha_{s}$ corrections. 
For the nucleon, there are two independent local interpolating operators: 
\begin{equation}
\eta _{1}(x)=\epsilon ^{abc}(u^{Ta}(x)C\gamma_{5} d^{b}(x))u^{c}(x),
\label{eq:eta1}
\end{equation}
\begin{equation}
\eta _{2}(x)=\epsilon ^{abc}(u^{Ta}(x)Cd^{b}(x))\gamma_{5} u^{c}(x). 
\label{eq:eta2}
\end{equation}
Here, $abc$ are color indices, $C$ stands for the charge conjugation matrix, while the spinor indices are omitted for simplicity. 
A general interpolating field can thus be expressed as 
\begin{equation}
\eta(x)= \eta_{1}(x)+\beta\eta_{2}(x),
\label{eq:interpolating field}
\end{equation}
where $\beta$ is a real parameter. 
From the time ordered correlator \cite{Ovchinnikov,KOhtani}, we can obtain the imaginary part of the old-fashioned
correlator 
of Eq.(\ref{eq:oldfash}) up to dimension 8 including first order $\alpha_s$ corrections. 
The details of  this calculation are explained in Appendix \ref{app:1}. 

To construct the final sum rule, we have to specify the kernel $W(q_{0})$. 
For finding its most useful form, we have tested several kinds of  $W(q_{0})$, such as 
$\exp(-\frac{q_{0}^{2}}{\mathrm{M}^{2}})$ and
$\frac{1}{\sqrt{4\pi \tau}}\exp(-\frac{(q_{0}^{2}-s)^{2}}{4\tau})$, which correspond to 
the Borel and Gaussian sum rules, respectively. 
Doing this, we however found that for these sum rules, the $\alpha_s$ corrections for the chiral even terms 
are very large and thus far from being under control. 
For example, the relative magnitude of the first order correction  
of the perturbative term 
does not depend on $\beta$ 
and reaches values as large as 90 \% of the
leading order term, when taking $\alpha_s = 0.5$. 
Furthermore, the Wilson coefficient of the four quark condensate $\langle \overline{q}q\overline{q}q\rangle$ 
of dimension 6 contains $\alpha_s$ corrections that can be even larger than the leading order term and seem to 
be converging only in a narrow region  around $\beta = -1$. These large corrections 
seriously put the validity of the above sum rules into question. 
Moreover, there is still another ambiguity present in the dimension 6 term, which comes from the 
approximation of reducing $\langle \overline{q}q\overline{q}q\rangle$ to $\langle \overline{q}q\rangle ^2 \kappa$, 
$\kappa$ being a dimensionless parameter. 
Although this value is known to be $1$ in the large $N_{c}$ limit 
\cite{Shifman1,Shifman2}, its value is only weakly constrained at $N_{c}=3$ \cite{Leinweber2}.

As we will explain below, it is possible to improve the above situation by choosing 
an appropriate kernel. To do this, we follow the method proposed by Ioffe and Zyablyuk \cite{Ioffe3}, 
who have constructed a new class of sum rules by using the phase rotated complex variable $q^2 e^{i\theta}$ 
instead of the real $q^2$. One advantage of this approach lies in the possibility of suppressing 
certain terms of the OPE by choosing some specific value of $\theta$. 

To apply this method to the parity projected sum rules, we change 
\begin{equation}
W(q_0)dq_{0} = \frac{1}{\sqrt{4\pi\tau}} q_0 \exp \Bigl(  -\frac{(q^2_0 - s)^2}{4\tau} \Bigr) dq_{0}
\end{equation}
to
\begin{equation}
\begin{split}
&W(q_0,\theta) dq_{0}=
\frac{1}{\sqrt{4\pi\tau}} \mathrm{Re} \Biggl[ q_0 e^{-i\theta} \\
& \quad \quad \quad \quad \quad \quad \times \exp \Bigl(  -\frac{(q^2_0 e^{-2i\theta} - s)^2}{4\tau} \Bigr) e^{-i\theta} dq_{0}\Biggr] ,
\end{split}
\label{eq:phaserokern}
\end{equation}
for which we note that the additional phase factor appearing in front of $dq_{0}$ is needed due to the 
change of the integration variable in Eq.(\ref{eq:finalpar}).  
Furthermore, let us make a comment on the analyticity of this kernel, which is essential for the 
formulation of the sum rules and thus might be a matter of concern as only the 
real part is retained in Eq.(\ref{eq:phaserokern}). However, as $q_0$ is real for the integrals in Eq.(\ref{eq:finalpar}), 
one can rewrite the kernel as
\begin{equation}
\begin{split}
&W(q_0,\theta) =
\frac{1}{\sqrt{4\pi\tau}} \frac{1}{2} \Biggl[ q_0 e^{-2i\theta} \exp \Bigl(  -\frac{(q^2_0 e^{-2i\theta} - s)^2}{4\tau} \Bigr) \\
&\quad \quad \quad \quad \quad \quad \quad + q_0 e^{2i\theta} \exp \Bigl(  -\frac{(q^2_0 e^{2i\theta} - s)^2}{4\tau} \Bigr)\Biggr],
\end{split}
\label{eq:phaserokern2}
\end{equation}
which of course is an analytic function and can thus be continued into the imaginary $q_0$ plane. As one further point, 
note that the kernel of Eq.(\ref{eq:phaserokern}) can only be used in a limited range of $\theta$, as for 
too large values of the phase, contributions from high-energy states to the sum rule will no more 
be exponentially suppressed. Specifically, one can easily convince oneself that $\theta$ 
should be chosen as $|\theta| < \frac{\pi}{8}$. 

Introducing the phase rotated kernel explained above has two essential advantages: 
1) By choosing an appropriate value for the phase $\theta$, one can largely reduce the $\alpha_s$ correction 
of the perturbative term and its relative contribution as a whole to the sum rule. 
2) Taking $W(q_0,\theta)$ to be an odd function of $q_0$, both the 
leading and a large part of the subleading $\alpha_s$ corrections of the dimension 6 term in the OPE, which are 
even functions of $q_0$, vanish. 
The contribution of this term to the sum rules is thus strongly suppressed, which means that the uncertainty 
originating from the value of the four-quark condensate has only a small influence on the results extracted 
from the sum rules. 

Substituting Eq.(\ref{eq:Pi1_old}) and Eq.(\ref{eq:phaserokern}) into Eq.(\ref{eq:finalpar}), 
we obtain  
\begin{equation}
\begin{split}
G^{\mathrm{old}\ \pm}_{\mathrm{OPE}}(s,\tau ) &\equiv 
\int_{-\infty }^{\infty}\frac{1}{\pi}\mathrm{Im}\Bigl [\Pi^{\ \pm }_{\mathrm{OPE}}(q_{0})\Bigr ] W(q_{0}, \theta) dq_{0} \\
&= \mathrm{G^{\mathrm{old}}_{1}}(s,\tau ) \pm \mathrm{G^{\mathrm{old}}_{2}}(s,\tau )\\
&= \int_{0}^{\infty} \rho ^{\pm}_{\mathrm{phys}}(q_{0}) W(q_{0}, \theta) dq_{0}.
\label{eq:newdef}
\end{split}
\end{equation}
Here, $\mathrm{G^{\mathrm{old}}_{1}}(s,\tau )$ and $\mathrm{G^{\mathrm{old}}_{2}}(s,\tau )$ are obtained as follows: 
\begin{equation}
\begin{split}
G^{\mathrm{old}}_{1}(s,\tau ) &= \int _{-\infty}^{\infty } \mathrm{Im}\biggl [\frac{\Pi ^{old}_{1-OPE}(q_{0})}{\pi} \biggr ]  
\cdot W(q_{0}, \theta) dq_{0} \\
&= (C_{0} + C_{0\alpha _{s}})\cos5\theta + C_{4}\langle \frac{\alpha _s}{\pi }G^{2}\rangle \cos \theta  \\
& \quad + (C_6 +C_{6\alpha _{s}}) \  \langle \overline{q}q\rangle ^2 \cos \theta \\
& \quad +C_{8}   \langle \overline{q}q\rangle  \langle \overline{q} g\sigma \cdot G q\rangle \cos3\theta ,
\label{eq:G1_complex}
\end{split}
\end{equation}

\begin{equation}
\begin{split}
G^{\mathrm{old}}_{2}(s,\tau ) &= \int _{-\infty}^{\infty } \mathrm{Im}\biggl [\frac{\Pi^{old}_{2-OPE}(q_{0})}{\pi} \biggr ] 
\cdot W(q_{0}, \theta) dq_{0} \\
          &= (C_{3} +C_{3\alpha _{s}} )\langle \overline{q}q\rangle \cos 2\theta 
                    + C_{5} \langle \overline{q}g\sigma \cdot Gq\rangle \\  
                    & \quad + C_{7} \langle \overline{q}q\rangle \langle \frac{\alpha _s}{\pi }G^{2} \rangle \cos 2\theta ,
\label{eq:G2_complex}
\end{split}
\end{equation}
where, $C_{0}$, $C_{0\alpha}$, $C_{3}$,$C_{3\alpha_{s}}$, $C_{4}$, $C_{5}$, $C_{6}$, $C_{6\alpha_s}$ and $C_{7}$ are given by 
\begin{equation}
\begin{split}
C_{0} &= \frac{5+2\beta +5\beta ^2 }{2^{11}\pi  ^4} I_{0} ,  \\
C_{0\alpha _{s}} &= \alpha_s \frac{5+2\beta +5\beta ^2 }{2^{11}\pi ^5} \\ 
&\quad \times \Bigl[\Bigl ( \frac{71}{12} + 2\theta \tan (5\theta )\Bigr )I_{0} 
- I_{0,\ln} \Bigr ],  \\
C_{3} &=- \frac{1}{2^{6}\pi  ^2}    
 \Bigl[  7 -2\beta -5\beta ^2   \Bigr ] I_{3} \\
C_{3\alpha _{s}} &=- \frac{\alpha_s}{2^{6}\pi  ^2}    
 \Bigl[  7\ \frac{15}{14} -2\beta \ \frac{3}{2} 
                 -5\beta ^2 \ \frac{9}{10}  \Bigr ] I_{3} \\
C_{4} &= \frac{5 +2\beta +5\beta ^2 }{2^{10}\pi ^2}  I_{4}, \\
C_{5} &= \frac{3(1-\beta ^{2})}{2^{6}\pi ^2}  I_{5}, \\
C_{6} &=0, \\
C_{6\alpha_{s}} &=\frac{\alpha_s}{2^{3}\ 3 \pi} \Bigl [ 7 \ \frac{47}{21} - 2\beta \ 
\frac{5}{3} -5\beta ^2 \ \frac{61}{15} \Bigr ] I_{6}, \\
C_{7}&=  -\frac{1}{\sqrt{4\pi \tau}} \frac{19 + 10\beta -29\beta ^2 }{2^{8}\ 3^{2}}  \mathrm{exp}(-\frac{s^2}{4\tau}), \\
C_{8}&=0.
\end{split}
\label{eq:OPEfres}
\end{equation}
Here, $I_n$ and $I_{n,\ln}$ are defined as 
\begin{equation}
\begin{split}
I_{n} &= \frac{1}{\sqrt{4\pi \tau}} \int _{0}^{\infty }  \mathrm{exp}\Bigl (-\frac{(q_{0}^{2}-s)^{2}}{4\tau} \Bigr ) \ q^{6-n}_{0} dq_{0},\\
I_{n,\ln}  &= \frac{1}{\sqrt{4\pi \tau}} \int _{0}^{\infty }  \mathrm{exp}\Bigl (-\frac{(q_{0}^{2}-s)^{2}}{4\tau}\Bigr ) \ q^{6-n}_{0} \ln ( q_{0}^{2} )dq_{0}.
\end{split}
\end{equation}
Note that $G^{\mathrm{old}}_{1}$ ($G^{\mathrm{old}}_{2}$) contains only terms of 
even (odd) dimensional condensates.  
In $G^{\mathrm{old}}_{1}$, the leading order Wilson coefficients of the 
dimension 6 and 8 terms such as  $\langle \overline{q}q\rangle^2$ and 
$\langle \overline{q}q\rangle  \langle \overline{q} g\sigma \cdot G q\rangle$ 
vanish due to the oddness of the kernel. 
For the same reason, the leading order terms of the purely 
gluonic condensates of dimension 6 and 8 ($\langle G^3 \rangle$, $\langle G^4 \rangle$) do not contribute to 
$G^{\mathrm{old}}_{1}$.  
We have therefore omitted 
these terms for simplicity.
For the technical details related to these statements, see Appendix A. 
\begin{table}[t]
\begin{center}
\begin{tabular}{|c|c|c|c|}
\hline
$\langle \overline{q}q\rangle $&$\langle \frac{\alpha _s}{\pi }G^{2}\rangle $&
$\langle \overline{q}g\sigma \cdot Gq\rangle $/$\langle \overline{q}q\rangle $
\\ \hline
$-(0.24\pm0.01)^{3} \,\mathrm{GeV}^{3}$&$0.012\pm 0.0036 \,\mathrm{GeV}^{4} $&$0.8\pm 0.2 \,\mathrm{GeV}^{2} $
\\ \hline
\end{tabular}
\caption{Values of the parameters appearing in the OPE, 
taken from \cite{Colangelo}.}
\label{the condensate parameters}
\end{center}
\end{table}
\begin{figure}[!tbp]
 \begin{center}
  \includegraphics[scale=0.68]{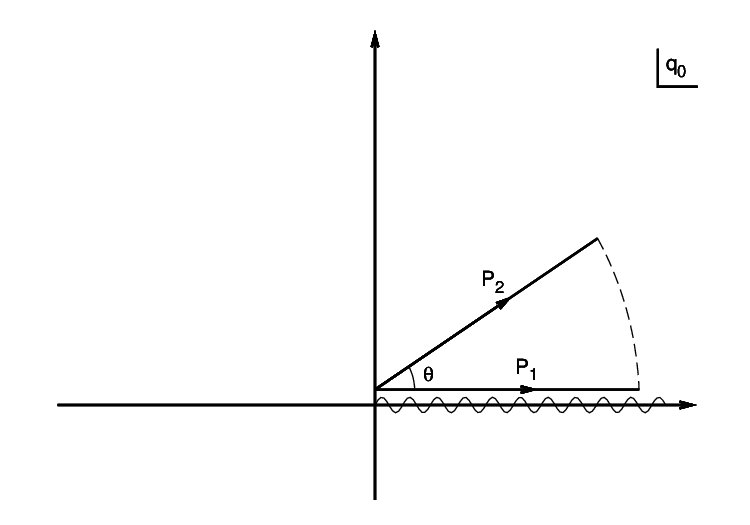}
 \end{center}
 \caption{The paths of $P_{1}$ and $P_{2}$, which are used for calculating the OPE result given in Eqs.(\ref{eq:G1_complex}-\ref{eq:OPEfres}). 
The wavy line denotes the non-analytic cut (or poles) of $\Pi^{\pm}(q_0)$ on the 
positive side of the real axis. 
}
 \label{fig:Contour_2}
\end{figure}
\begin{figure}[!tbp]
 \begin{center}
  \includegraphics[scale=0.68]{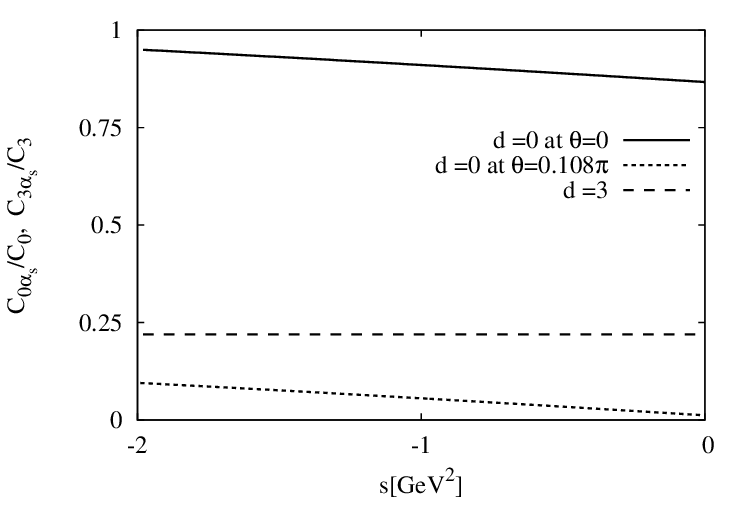}
 \end{center}
 \caption{The ratios of $\alpha_{s}$ corrections to leading order terms of dimension 0 and 3: $\frac{C_{0\alpha_s}}{C_{0}}(\theta)$ and $\frac{C_{3\alpha_s}}{C_{3}}(\theta)$ 
at $\beta=-0.9$. These values are shown as a function of $s$, with $\tau$ fixed to $0.5\,\textrm{GeV}^4$. 
The solid and dotted lines shows $\frac{C_{0\alpha_s}}{C_{0}}$ at $\theta = 0.108\pi$ and $\theta=0$, respectively. 
The dashed line gives $\frac{C_{3\alpha_s}}{C_{3}}$, which does not depend on $\theta$.
}
 \label{fig:OPE_ratio}
\end{figure}
\begin{figure*}[!tbp]
 \begin{center}
  \includegraphics[scale=1.25]{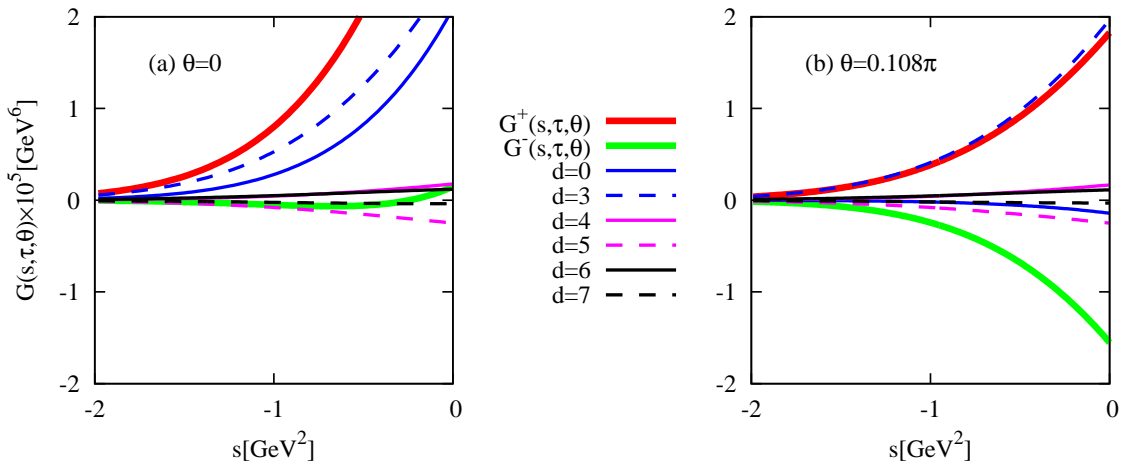}
 \end{center}
 \caption{(a) $G^{\mathrm{old}\ \pm}_{\mathrm{OPE}}(s,\tau )$ and  
the respective OPE contributions from dimension 0 to 7 at $\tau = 0.5[\mathrm{GeV}^{4}]$, $\beta=-0.9$ and $\theta=0$. 
 The thick solid lines denote $G^{\mathrm{old}\ \pm}_{\mathrm{OPE}}(s,\tau )$ 
and the thin solid (dashed) lines the chiral even (odd) OPE terms. (b) Same as for (a), but for $\theta=0.108\pi$.
}
 \label{fig:OPE data_0 and theta}
\end{figure*}

The values of the vacuum condensates appearing in the above equations are shown in 
Table. \ref{the condensate parameters}. For all the above results, the renormalization constant $\mu$ has been set to 1 GeV. 
Related to this choice, we use the value of $\alpha_{s} = 0.5$ which corresponds to the energy scale of 1 GeV \cite{Bethke}. 
To obtain the concrete form of $G^{\mathrm{old}}_{1}(s,\tau )$ and $G^{\mathrm{old}}_{2}(s,\tau )$, we have made use the fact that the contributions of the path $P_{1}$ 
are as same as those of $P_{2}$ in Fig. \ref{fig:Contour_2} due to the Cauchy theorem. 

To determine a suitable value of $\theta$ for which the $\alpha_s$ corrections of 
the perturbative term are reduced, 
we find that $\frac{71}{12}$ and $2\theta \tan(5\theta ) $ in $C_{0\alpha_{s}}$ cancel each other out at around $\theta=0.108\pi$, 
while $I_{\ln}$ gives only a small contribution. 
Hence we set the value to $0.108\pi$ throughout this work. To illustrate the effect of the introduced phase, the 
ratios of $\alpha_{s}$ corrections to leading terms at dimension 0 and 3: $\frac{C_{0\alpha_s}}{C_{0}}(\theta=0.108\pi)$, $\frac{C_{0\alpha_s}}{C_{0}}(\theta=0)$ and $\frac{C_{3\alpha_s}}{C_{3}}$ 
(which does not depend on $\theta$) are shown in Fig. \ref{fig:OPE_ratio} for $\tau = 0.5\, \textrm{GeV}^4$. 
As can be observed from this figure, the contribution of the $\alpha_s$ term of dimension 0 is reduced to $5\,\%$ of the 
leading term, which shows that the convergence of the perturbative expansion is significantly improved. 

The contributions of each term to $G^{\pm}(s,\tau)$ at $\theta=0$ and $\theta=0.108\pi$ are given in 
Figs. \ref{fig:OPE data_0 and theta} (a) and (b), respectively. It firstly should be noted here that for both 
cases, the contribution of dimension 6 is small and that the uncertainty of the four-quark condensate 
therefore does not lead to much ambiguities for these sum rules. This is caused by our use of 
an $q_0$-odd kernel, which eliminated the leading order contribution of dimension 6. 
For $\theta=0$, the dimension 0 and 3 terms are dominant, 
which means that not only low lying nucleon states 
but also the continuum (which mainly originates from the perturbative dimension 0 term) largely contributes to $G^{\pm}(s,\tau)$. 
This situation makes it difficult to extract information on nucleonic properties. 
Especially, for the negative parity states, the extraction is highly complicated due to the strong cancellation between the terms of dimension 0 and 3. 
On the other hand, 
from Fig. \ref{fig:OPE data_0 and theta} (b) it can be seen 
that in the phase rotated sum rule the chiral condensate term clearly gives the dominant contribution to the OPE. 
Therefore, unlike the sum rule at $\theta =0$, the disturbance due to the contribution of the continuum 
is considerably reduced. 
In Fig. \ref{fig:OPE data_0 and theta} (b), it can also be observed that the difference between the OPE for positive and negative 
parity is mainly caused by the switched sign of the quark condensate term. 
Since the spontaneous breaking of the chiral symmetry is considered to cause the mass splitting of the positive and negative parity
states of the nucleon, this sum rule is likely to be useful for investigating 
the relation between the chiral symmetry and the mass splitting between positive and negative parity nucleons, which we plan to investigate in detail in a future work. 

\section{The maximum entropy method}
\label{sec:3} 
In this section, we briefly introduce the maximum entropy method (MEM), 
and explain how this approach applies to the analysis of QCD sum rules. 
The equation to be analyzed can be written down as
\begin{equation}
G_{\mathrm{OPE}}(x)=\int^\infty _{0} W(x,q_{0} )\rho (q_{0}) dq_{0}. 
\label{eq:sr} 
\end{equation}
Here, $x$ stands for the parameters occurring in the kernel W($x$,$q_{0}$), such as s and $\tau$. 
In a QCD sum rule analysis, one aims to extract information on the spectral function 
from $G_{\mathrm{OPE}}(x)$. 
However, this function is only known with limited precision due to the uncertainties of the values of the 
various vacuum condensates. 
Furthermore, only the limited region of $x$, 
where the OPE converges, can be used for the analysis. 
Therefore, rigorously solving Eq.(\ref{eq:sr}) is an ill-posed problem.
Nevertheless, the MEM technique enables us at least to statistically 
determine the most probable form of $\rho(q_{0})$.

For this purpose, we define 
$P[\rho |GH]$, which is the conditional probability of $\rho$ given $G_{\mathrm{OPE}}$ and H, 
H representing additional knowledge on the spectral function such as positivity and asymptotic values. 
Using Bayes' theorem, $P[\rho |GH]$ is rewritten as 
\begin{equation}
P[\rho |GH]=\frac{P[G|\rho H]P[\rho |H]}{P[G|H]},
\label{eq:MEMdummy1}
\end{equation}
where 
$P[\rho |H]$ is the so-called prior probability, and $P[G|\rho H]$ stands for the 
likelihood function.
The most probable form of $\rho(\omega)$ is obtained, by maximizing $P[\rho |GH]$ as given in Eq.(\ref{eq:MEMdummy1}). 
For carrying out this task, one needs the concrete functional forms of 
$P[\rho |H]$ and $P[G|\rho H]$, which can be expressed as \cite{Gubler1,Asakawa}   
\begin{equation}
\begin{split}
P[G|\rho H]&\propto e^{-L[\rho ]},\\
P[\rho |H]&\propto e^{\alpha S[\rho ]}. 
\end{split}
\end{equation}
Here, $\alpha$ is a real positive number, which determines the relative weight between $L[\rho ]$ and $S[\rho ]$. 
The functional $L[\rho]$ is normally used for $\chi ^{2}$ fitting and can be given as 
\begin{equation}
\begin{split}
L[\rho] =&\frac{1}{2(x_{\mathrm{max}}-x_{\mathrm{min}})}\int^{x_{\mathrm{max}}}_{x_{\mathrm{min}}}dx \,\, \times \\
&\quad \quad \frac{[G_{\mathrm{OPE}}(x)-\int^\infty _{0} W(x,q_{0} )\rho (q_{0}) dq_{0}]^2}{\sigma ^{2}(x)}.
\label{eq:likelihood cont}
\end{split}
\end{equation}
The error $\sigma(x)$ in the above equation is determined from the uncertainty of the vacuum condensates 
and is evaluated by the statistical method explained in \cite{Leinweber2}. 
$S[\rho]$ is known as the Shannon-Jaynes entropy and can be obtained as 
\begin{equation}
S[\rho] =\int^{\infty}_{0} dq_{0} [\rho  (q_{0} ) -m(q_{0} ) - \rho (q_{0} ) 
\log(\frac{\rho (q_{0} )}{m(q_{0} )})],
\label{eq:entropy cont}
\end{equation}
where the function $m(\omega)$ is called the default model and is an input of the MEM method. 
It namely corresponds to the function which maximizes $P[\rho |GH]$ if no information from the OPE is available (or 
its uncertainty is infinitely large). We therefore use $m(\omega)$ to specify the values 
of the spectral function at very high and very low energy, for which $G_{\mathrm{OPE}}(x)$ provides only 
a weak constraint. 

Since $P[G|H]$ does not depend on $\rho (\omega )$, 
it can be regarded as a simple normalization constant and we can thus ignore it. 
From the preceding equations, we now have the specific form of $P[\rho |GH]$:
\begin{equation}
\begin{split}
P[\rho |GH]&\propto P[G|\rho  H]P[\rho |H]\nonumber\\
 &= e^{Q[\rho ]},
\end{split}
\end{equation}
where 
\begin{equation}
Q[\rho ]= \alpha S[\rho ]-L[\rho ].              
\end{equation}
Therefore,  to get the most probable $\rho(\omega)$, 
we have to solve the numerical problem of obtaining the form of $\rho(\omega)$ that maximizes Q[$\rho $].

As for the free parameter $\alpha$, we treat it in accordance with 
the conventional Bryan algorithm, which means that it is integrated out at the end of the calculation (see \cite{Bryan} for details). 
Using this approach, we can get the most probable spectral 
function and additionally its error $\langle \delta \rho \rangle$, averaged in some specific 
energy region. 
In the figures of this paper, we show the errors as three horizontal lines, whose lengths 
stand for the region of $q_{0}$, for which the error $\langle \delta \rho \rangle$ is calculated, while their heights correspond to 
$\langle \rho \rangle + \langle \delta \rho \rangle$, 
$\langle \rho \rangle$, 
$\langle \rho \rangle - \langle \delta \rho \rangle$, 
respectively, $\langle \rho \rangle$ representing the value of the spectral function averaged over the region specified 
by the horizontal length of the error bars. 
For more details about MEM, we refer the reader to \cite{Asakawa,Jarrel}. 

\section{Analysis of mock and OPE data}
\label{sec:4} 
Firstly, let us explain how the default model $m(q_{0})$, the analyzed parameter regions of 
$\tau$ and $s$ and the values of $\beta$ are determined. 

As mentioned in the last section, $m(q_{0})$ should reflect our prior knowledge of the spectral function 
such as the asymptotic behavior in the high or low energy region. 
Hence one may naturally conceive three types of default models.
The first one is a constant consistent only 
with the asymptotic behavior of the spectral function at low energy (which is close to zero as the spectral function 
does not have any strength below the nucleon ground state), the second one is also a constant 
which reflects the asymptotic behavior at high energy (which is obtained perturbatively) 
and the third one is a combination of the first and second with the correct behavior at both 
high and low energy, which we call as the hybrid default model.
Carrying out the analyses with three types default models, 
we have confirmed that the position of the lowest peak in the extracted spectral function does not much depend on this choice. 
Therefore, since only the hybrid model reflects the correct behavior of the spectral function 
at both high and low energy, 
we will in this paper only show the results using this default model. 
Its specific form is parametrized as follows: 
\begin{equation}
\begin{split}
m_{\mathrm{hybr}}(\omega )&=\frac{5+2\beta +5\beta ^2 }{128(2\pi ) ^4} 
\frac{1}{1+e^{(q_{\mathrm{th}} -q_{0} )/\delta}}.
\label{eq:default hybr}
\end{split}
\end{equation}
The concrete values of $q_{\mathrm{th}}$ and $\delta$ will be determined in the next subsection. 

Next, we discuss the employed parameter region of $\tau$ and $s$, which is mainly 
restricted by the convergence of the OPE. 
Since the calculation of the correlation function using the OPE is truncated at a certain order, 
one should keep the contribution of the highest dimensional term small to justify the truncation and to reduce the possibility that higher order terms spoil the result. 
Therefore, we use the well established criterion that the ratio of the highest dimensional term 
is less than 0.1 of the whole $G_{\mathrm{OPE}}(x)$. 
From the property of the kernels, we expect that 
the spectral function at small values of $\tau$ will be more sensitive to 
narrow structures such as the lowest peak, 
while the spectral function at larger values will to a large extent be fixed by the continuum. 
\begin{table}[t!]
\begin{center}
\begin{tabular}{|c|c|c|c|c|}
\hline 
$\tau$                                &0.5  &1.0  &1.5  &2.0  \\ \hline
$s_{\mathrm{min}}$ of positive parity &-1.91&-4.32&-6.75&-9.19 \\ \hline
$s_{\mathrm{min}}$ of negative parity &-1.22&-3.10&-5.02&-6.94 \\
\hline
\end{tabular}
\caption{Values of $s_{\mathrm{min}}$ [GeV$^2$] at $\beta=-0.9$ and fixed $\tau$ [GeV$^4$].}
\label{tab:s_{min} at beta=-0.9}
\end{center}
\end{table}
Hence, to extract as much information as possible from $G_{\mathrm{OPE}} (x )$, 
we use several values of $\tau$ at the same time ($\tau$ = 0.5, 1, 1.5, 2$\,\mathrm{GeV}^4$) and determine the lowest value of $s$ (denoted as $s_{\mathrm{min}}$) at each $\tau$. 
The maximum values of $s$ (denoted as $s_{\mathrm{max}}$), are set to $s_{\mathrm{max}} = s_{\mathrm{min}} + 1\,\mathrm{GeV}^2$. 
For the detailed procedure of the MEM analysis using simultaneously two adjustable parameters, 
we refer the reader to \cite{KOhtani}. 

As for the value of $\beta$ in Eq.(\ref{eq:interpolating field}), we find that the most reliable results can be obtained around $\beta=-0.9$.
For other values of $\beta$ such as 0 and $\infty$, 
the contribution of the mixed condensate to the OPE becomes large and thus the MEM analysis 
only leads to spectral functions consistent with the input default model. This happens because of 
the large uncertainty of the respective condensate value, which weakens the constraint of the likelihood function of Eq.(\ref{eq:likelihood cont}).  
Near $\beta = 1$, the subleading dimension 6 term proportional to $\langle \overline{q}q\rangle ^{2}$ turns out to be 
the largest contribution to the OPE. Because of the large uncertainty related to this term, this is not a very useful sum rule. 
Therefore, we carry out the analyses at $\beta=-0.9$. 
The respective minimum values of $s$ at each $\tau$ determined by the above criterion 
are given in Table. \ref{tab:s_{min} at beta=-0.9}. 

Finally, let us briefly make a comment on 
the high dimensionality of the spectral function for baryonic channels. 
It is easily understood from dimensional considerations that unlike in the meson case, 
the contribution of the continuum states to the 
baryon spectral function is proportional to $q_{0}^5$ and thus 
strongly enhanced. 
As was done in similar studies using MEM and lattice QCD, 
we will analyze $\rho(q_{0})$/$q_{0}^5$ instead of $\rho(q_{0})$. 
we denote $\rho(q_{0})$/$q_{0}^5$ as $\rho(q_{0} )$ from now on, meaning that Eq.(\ref{eq:sr}) changes to: 
\begin{equation}
\begin{split}
G_{\mathrm{OPE}}(x)=\int^\infty _0 W(x,q_{0} )q_{0}^{5} \ \rho(q_{0}) dq_{0}.
\end{split}
\end{equation}

\subsection{Analyses using mock data}
\label{subsec:4.1} 
\begin{figure*}[tbp]
 \begin{center}
  \includegraphics[width=15cm]{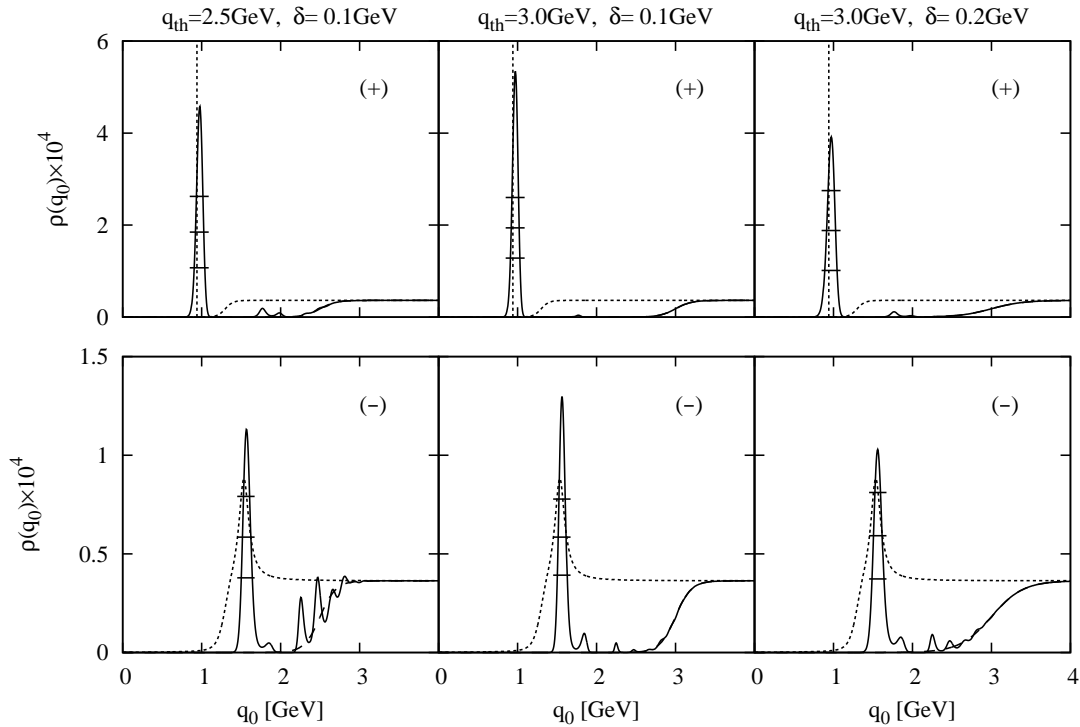}
 \end{center}
 \caption{Spectral functions extracted from the MEM analyses of $G_{\mathrm{mock}}(x)$ 
using various default models at $\beta = - 0.9$. 
The solid, dotted and dashed lines show $\rho(q_{0})$, the mock data of Eqs.(\ref{eq:mockspecp}) and (\ref{eq:mockspecn}) and the default model of Eq.(\ref{eq:default hybr}) respectively. 
Except for the lower left figure, the default model cannot be seen since it almost completely overlaps with the emerging continuum. 
The values of parameters employed for the default model are given on top of the figure at each row. 
The parity of the corresponding spectral function is shown on the top right corner of each figure.
}
 \label{fig:Analysis_mock}
\end{figure*}

In order to clarify the effectiveness of the MEM analysis and 
to obtain suitable values for the parameters determining the default model, 
we first analyze the mock data constructed from a phenomenological spectral function.
The specific expressions of the positive and negative parity mock data are 
\begin{equation}
\begin{split}
\rho ^{+}_{\mathrm{mock}}(q_{0} )=\frac{\lambda _{L}^{+2}}{M_{L} ^{+5}}\delta (q_{0} -  M_{L}^+) +  
\frac{5+2\beta +5\beta ^2 }{128(2\pi ) ^4}
\frac{1}{1+ e^{\frac{(\omega -q_{0} )}{\delta}}} 
\label{eq:mockspecp}
\end{split}
\end{equation}
and 
\begin{equation}
\begin{split}
\rho ^{-}_{\mathrm{mock}}(q_{0} )=&\frac{\lambda _{L}^{-2}}{M_{L} ^{-5}} \frac{\Gamma}{2\pi} \frac{1}{\bigl( q_0- M_{L}^- \bigr) ^2 + \Gamma ^2 /4 } \\ 
&\hspace{2.0cm} + \frac{5+2\beta +5\beta ^2 }{128(2\pi ) ^4}
\frac{1}{1+ e^{\frac{(\omega -q_{0} )}{\delta}}}, 
\label{eq:mockspecn}
\end{split}
\end{equation}
respectively. 
These mock data are shown in  Fig. \ref{fig:Analysis_mock} as dotted lines. 
Here, we use $\omega=1.3$ GeV and $\delta=0.05$GeV. 
For the positive (negative) parity states, the mass of the lowest lying state $M_{L}^{\pm}$ and the residue $\lambda ^{\pm 2}_{L}$ are taken as 
$M_{L}^{\pm}=0.94 \ (1.535)$ GeV, $\lambda ^{\pm 2}_{L}=5.0\times 10^{-5} \ (1.0\times 10^{-4}) \mathrm{GeV}^6$. 
For the negative parity channel, to reflect the finite width of the peak, we use the Breit-Wigner form and set its width to $\Gamma =0.15$ GeV, which is the 
corresponding PDG value of the N(1535). 

We apply the MEM analysis to 
$G_{\mathrm{mock}}(x) \equiv \int^\infty _0dq_{0} W(x,q_{0})\rho _{\mathrm{mock}} (q_{0} )$ 
instead of $G_{OPE}(x)$. 
For the error entering Eq.(\ref{eq:likelihood cont}), we use the one obtained from the OPE data to mimic the actual analysis to be carried out later. 
The results are shown in Fig. \ref{fig:Analysis_mock} as solid lines. 
In all cases, the peak position is successfully reconstructed at about the input value $M_{L}^{\pm}$, 
the differences between $M_{L}^{\pm}$ and the position of the obtained peaks being less than 40 MeV. 
Furthermore, it it seen that the details of the default model do not alter the position of the lowest peak, which is our main object of interest. 
Hence, there exist several default models which are equally valid for the problem studied here.
Our choice for the following analysis of 
OPE data will be $q_{th}=3.0\,\mathrm{GeV}$ and $\delta =0.1\,\mathrm{GeV}$ for both the positive and negative spectral 
functions.

In contrast to the peak position, we observe that the width of the lowest negative parity state and the continuum for both parities are not reproduced. 
The issue of the width was discussed in \cite{Gubler1} for the Borel kernel, where it was pointed out that 
the extracted peak widths are mainly caused by the limited resolution of the MEM procedure and thus should be 
considered to be an artifact of the present method. 
As for the continuum, it is not reconstructed well and its behavior strongly depends on the choices of the default model. 
This behavior can be understood by considering the properties of  $W(x,q_{0})$.  
Firstly, one notes that 
$W(x,q_{0})$ of Eq.(\ref{eq:phaserokern}) oscillates with increasing frequency as 
$q_{0}$ becomes larger. 
Therefore, in the high $q_0$ region, 
the contribution of the continuum is largely reduced due to the fast oscillating kernel, 
which makes it difficult to reproduce its properties.  
The relatively narrow width of the obtained negative parity peak can also be explained by the oscillation kernel, which suppresses the tail region of the peak. 
Finally, let us compare the results obtained from MEM with the ``pole + continuum" model. 
Using MEM, we have derived the spectral function without the ``pole + continuum" ansatz. 
However, the obtained spectral functions have peaks with a narrow width and a continuum which essentially only depends 
on the default model. 
Therefore, for the phase rotated sum rule, we conclude that MEM does not extract more information on the 
lowest state than an analysis employing the ``pole + continuum" ansatz. 
 
\subsection{Analyses using OPE data}
\label{subsec:4.2} 
\begin{figure*}[!tbp]
 \begin{center}
  \includegraphics[width=11.5cm]{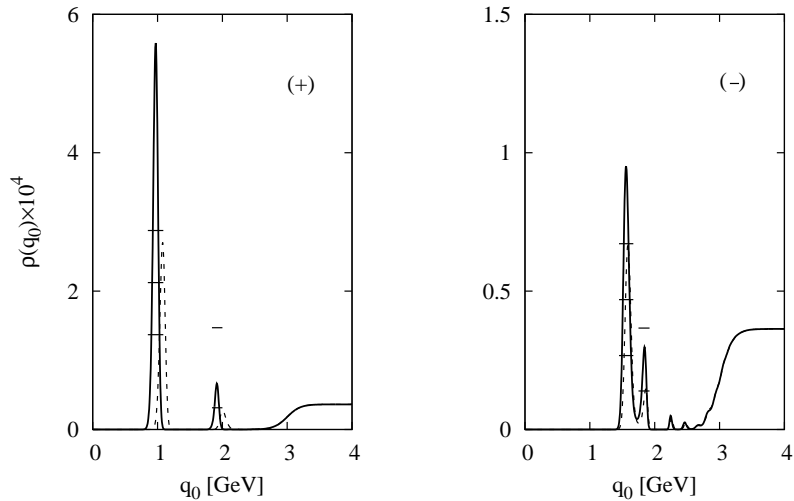}
 \end{center}
 \caption{The positive (left) and negative parity (right) spectral functions extracted from MEM analyses of the OPE data $G^{\mathrm{old}\ \pm}_{\mathrm{OPE}}(s,\tau )$
of Eqs.(\ref{eq:G1_complex}-\ref{eq:OPEfres}) at $\beta$ = $-0.9$ (solid lines). 
The dashed lines show the respective results for omitted $\alpha_s$ corrections. 
The parity of the corresponding spectral functions is shown on 
the top right corner of each figure. 
}
 \label{fig:Analysis_OPE_complex}
\end{figure*}

Having finished all necessary preparations, we can now carry out the analysis using the real OPE data 
$G^{\mathrm{old}\ \pm}_{\mathrm{OPE}}(s,\tau )$, given in Eq.(\ref{eq:OPEfres}). 
The corresponding spectral functions are shown in Fig. \ref{fig:Analysis_OPE_complex}, 
where the results with (without) $\alpha_{s}$ corrections are shown as solid (dashed) lines. 
Comparing the solid and dashed lines, it firstly 
should be noted that the qualitative behavior of the spectral functions does not change much, 
which shows that the nucleon properties are mainly determined by the non-perturbative condensates (which 
is essentially the chiral condensate of dimension 3 here), and not by the perturbative corrections. 
On the other hand, the lowest peaks derived from the full OPE of Eq.(\ref{eq:OPEfres}) are sharper and closer to the experimental values than 
those including only the leading order Wilson coefficients. 
These findings indicate that the inclusion of the $\alpha_s$ corrections improves the accuracy of the sum rules. 
Furthermore, we can observe that the $\alpha_s$ terms cause an increase of the 
residue of the ground state, which is qualitatively consistent with the analysis presented in \cite{Sadovnikova}.

Let us now discuss in detail the results of the analysis using the full Eq.(\ref{eq:OPEfres}). 
As could be expected from the OPE data shown in Fig. \ref{fig:OPE data_0 and theta} (a), the positive and 
negative parity spectral functions behave differently and the splitting between the two lowest peaks is 
clearly observed. 
In the positive parity spectral function, peaks are found at 950 MeV and 1930 MeV.  
As can be inferred from the error bars, the lowest peak which corresponds to the nucleon ground state is 
statistically significant, while the second one is not. 
Furthermore, it is observed that while we can extract a clear signal of the ground state, no indication for the first excited positive parity 
state (the Roper, N(1440)) is extracted, which most likely means that this state only couples weakly to our employed interpolating field. 

For negative parity, two peaks appear at 1540 MeV and 1840 MeV. 
The lowest peak appears close to the experimentally observed lowest negative parity state, the N(1535). 
However, it has to be emphasized that we can not conclude that the lowest 
peak solely corresponds to the N(1535), because of the possible contribution of the next lying N(1650) state. 
As we have explained in the last section, the analysis used here does not have the ability to reconstruct 
the peak width correctly. 
As elucidated in \cite{KSuzuki}, it is hence quite difficult to reproduce two separated peaks 
lying in a narrow energy region. 
Therefore, the only firm conclusion to be drawn from our analysis 
is that some negative parity state exists near 1540 MeV. 
We furthermore find a second peak in the negative parity spectral function, its existence is, however, much less certain 
because of the large errors involved. We thus do not speculate about its possible physical 
interpretation here. 

\section{Summary and conclusion}
\label{sec:5}  
We have analyzed the nucleon spectral function using QCD sum rules and the 
maximum entropy method. 
The first order $\alpha_{s}$ corrections are taken into account along with the parity projection. 
Analyzing the parity projected nucleon sum rule, we have faced several problems. 
Firstly, there were some issues of technical nature in the approach proposed in \cite{Jido}, 
leading to potential ambiguities in the higher order terms of the OPE. Secondly, and most severely, employing 
conventionally used kernels (Borel, Gaussian), we found 
that the
$\alpha_{s}$ corrections tend to be too large for the perturbative expansion of the Wilson coefficients to converge, 
which considerably lowers the
reliability of the analysis. 
Moreover, the continuum gives a large contribution to these sum rules, which 
makes the extraction of the information on nucleonic states difficult. 

To solve these problems, we have 
firstly improved the parity projected sum rules using the old-fashioned correlation function 
and have, by this, placed the approach of \cite{Jido} on a mathematically solid ground. 
We then have constructed the parity projected sum rule for the nucleon 
including all known first order $\alpha_{s}$ corrections. 
Furthermore, by using the phase-rotated Gaussian kernel, we could suppress the $\alpha_{s}$ corrections 
and could thus improve the convergence of the corresponding perturbative expansion 
and, at the same time, could reduce the continuum contribution. 
Moreover, we have applied the MEM analysis to this sum rule. 

In order to check the effectivity of the MEM technique, we have first carried out 
analyses using mock data and have confirmed that the positions of the lowest peak of the spectral function can be reliably extracted without strong dependence 
on the default model. 
On the other hand, the obtained peaks all have a relatively 
narrow width, which does not much depend on the input mock data. Furthermore, the generated continuum is essentially determined by the default model. 
These facts indicate that MEM is not able to extract precise information on the peak width and the continuum threshold from the OPE data. 
Therefore, for the phase-rotated kernel used in this paper, MEM leads to a result which would presumably not differ much from the outcome of a similar analysis 
using the ``pole + continuum" ansatz. 
Next, we have analyzed the OPE data, obtained the spectral functions of both positive and negative 
parity and found that both  spectral functions contain significant peaks near the experimental values of the lowest lying states. 
As the term proportional to the dimension 3 quark condensate, which switches its sign depending on the parity, 
dominates the OPE used here, this sum rule result provides novel evidence for the scenario in which the spontaneous 
breaking of chiral symmetry causes the splitting between positive and negative parity baryon states. 

Related to the subjects studied in this paper, there are, however, still a few unresolved problems left.  
One such issue is the Roper resonance N(1440), 
which we are not able to find in the present analysis. 
One possible reason for our non-observation of the Roper is that in 
this study, we can only use an interpolating field operator with $\beta =-0.9$ to obtain reliable results. 
Hence, we cannot exclude the possibility that the N(1440) could be observed at other values of $\beta$. 
For such an analysis to become feasible, it would first of all be necessary to determine the dimension 5 mixed condensate 
and the dimension 6 four-quark condensate 
with higher precision, which could be possible in future lattice QCD studies. 
For the negative parity states, we presently cannot determine whether the 
obtained peak corresponds to either N(1535) or N(1650), or both of them. 
This is due to the limited resolution of the
MEM analysis, which is closely related to the error of
the OPE data which comes from the uncertainty of the
various vacuum condensates. 
Therefore, if the values of these condensates are determined more accurately in the future, 
a more detailed analysis of the nucleon spectral functions of both parities may become possible. 

\begin{acknowledgments}
This work is partially supported by KAKENHI under Contract Nos. 19540275, 22105503 and 24540294. 
P.G. gratefully acknowledges the support by the Japan Society for the 
Promotion of Science for Young Scientists (Contract No. 21.8079). 
K. O also acknowledges the financial support from the Global
Center of Excellence Program by MEXT, Japan through the
"Nanoscience and Quantum Physics" Project of the Tokyo Institute of Technology.
The numerical calculations of this study have been partly performed on the super grid 
computing system TSUBAME at Tokyo Institute of Technology.
\end{acknowledgments}

\appendix
\section{Construction of the sum rule from the old-fashioned correlation function}
\label{app:1}
In this appendix, we discuss the details of the calculations for obtaining general parity projected baryonic QCD sum rules. 
\subsection{Derivation of the old-fashioned correlation function}
We derive in this subsection the old-fashioned correlation function from the time ordered correlator. 
In the course of this calculation, we will repeatedly make use of 
the Fourier transform of the Heaviside step-function: 
\begin{equation}
\begin{split}
 \theta (x_{0})=\frac{1}{2\pi i} \int d k_0 \frac{1}{k_0 - i\epsilon} e^{ik_0x_0}. 
\label{eq:F_step}
\end{split}
\end{equation}
Firstly, let us consider the phenomenological side of the old-fashioned correlator, 
which can be calculated as follows: 
\begin{equation}
\begin{split}
\Pi^{\mathrm{old}}(q_0) &= i \int d^{4}x e^{iqx} \theta(x_0) \langle0|T[\eta(x)\overline{\eta}(0)]|0\rangle\Big|_{\vec{q}=0} \\
&=\displaystyle  - \int d^4x e ^{iqx}\theta(x_{0}) \int \frac{d^4p}{(2\pi )^{4}} e ^{-ipx}  \int _{0}^{\infty} dm \ \times \\
& \biggl[ \rho_{+}(m) \frac{\slash{p} + m}{p^2 -m^2 +i\epsilon} + \rho_{-}(m) \frac{\slash{p} - m}{p^2 -m^2 +i\epsilon}\biggr] \Bigg|_{\vec{q}=0} \\
&= - \frac{1}{2\pi i} \int d k_0 \frac{1}{k_0 - i\epsilon} \displaystyle  \int _{0}^{\infty}dm \\
&\quad \biggl[  \rho_{+}(m) \frac{(k_0 + q_0)\gamma_0 - \vec{q} \cdot \vec{\gamma} + m}{(k_0 + q_0)^2 - \vec{q}^2 -m^2 +i\epsilon} \\
& \quad + \rho_{-}(m)
\frac{(k_0 + q_0)\gamma_0 - \vec{q} \cdot \vec{\gamma} - m}{(k_0 + q_0)^2 - \vec{q}^2 -m^2 +i\epsilon} 
\biggr]  \Bigg|_{\vec{q}=0}.
\end{split}
\label{eq:forward_phys1}
\end{equation}
Let us here briefly explain the difference between this paper and the previous study of \cite{Jido}. 
We have here replaced $\eta(x)\overline{\eta}(0)$ by $T[\eta(x)\overline{\eta}(0)]$, where T stands for the time ordered product. 
This change is implemented because, strictly speaking, it is only the T-product, for which perturbative 
calculations using Feynman diagram techniques are allowed. 

The integrand of the above result contains three poles, two in the upper half of the imaginary plane of $k_0$ and 
one in the lower half. Therefore, closing the contour of integration of $k_0$ in the lower half and using the Cauchy theorem, 
we pick up the residue of one pole and get 
\begin{equation}
\begin{split}
&\Pi^{\mathrm{old}}(q_0) 
= - \frac{1}{2} \displaystyle \int _{0}^{\infty} dm \times  \\
&\hspace{0.5cm}\biggl[ \rho_{+}(m) \frac{1}{q_0 - \sqrt{\vec{q}^2 + m^2} + i\epsilon} 
\big[ \gamma_0 - \frac{\vec{q} \cdot \vec{\gamma} - m}{\sqrt{\vec{q}^2 + m^2} - i\epsilon} \big] + \\
& \hspace{0.3cm} \rho_{-}(m)  \frac{1}{q_0 - \sqrt{\vec{q}^2 + m^2} + i\epsilon} 
\big[ \gamma_0 - \frac{\vec{q} \cdot \vec{\gamma} + m}{\sqrt{\vec{q}^2 + m^2} - i\epsilon} \big] \biggr]
\Bigg|_{\vec{q}=0} \\
&=  - \frac{1}{2} \displaystyle \int _{0}^{\infty} dm \biggl[ \rho_{+}(m) \frac{1}{q_0 - m + i\epsilon}(\gamma_0 + 1)  \\
& \hspace{2.5cm} +\rho_{-}(m) \frac{1}{q_0 - m + i\epsilon}(\gamma_0 - 1) \biggr].
\end{split}
\label{eq:forward_phys2}
\end{equation}
From this result, it can be understood that the 
old-fashioned correlator contains only poles in the positive $q_0$ 
region and is analytic for $\mathrm{Im}\,q_0 \geq  0$. 
Furthermore, by applying the projection operator $\frac{1}{2}(\gamma_0 \pm 1) \equiv P^{\pm}$ to the above equation 
and taking the trace over the spinor indices, we can obtain Eq.(\ref{eq:Im}): 
\begin{equation}
\begin{split}
&\frac{1}{\pi}\mathrm{ Im} \Bigl[ \frac{1}{2} \mathrm{Tr} \big[P^{\pm} \Pi^{\mathrm{old}}(q_0)\big] \Bigr] \\
&=\frac{1}{\pi}\mathrm{ Im} \Bigl[ - \displaystyle \int _{0}^{\infty} \rho_{\pm}(m) \frac{1}{q_0 - m + i\epsilon} dm \Bigr ] \\
&= \displaystyle \rho_{\pm}(q_{0})
\end{split}
\label{eq:forward_phys3}
\end{equation}

Next, we consider the OPE side of the old-fashioned correlator.
For a local interpolating field constructed from three light quarks, 
the time ordered correlation function in coordinate space can generally be expressed as 
\begin{equation}
\begin{split}
\Pi(x) =&
\big[ C^{(0)}_x \frac{1}{(x^2 - i\epsilon)^5} + C^{(4)}_x \frac{1}{(x^2 - i\epsilon)^3} \\
&+ C^{(6)}_x \frac{1}{(x^2 - i\epsilon)^2} 
+ C^{(8)}_x \frac{1}{x^2 - i\epsilon} + \dots \big]\slash{x} \\
&+ C^{(3)}_x \frac{i}{(x^2 - i\epsilon)^3} + C^{(5)}_x \frac{i}{(x^2 - i\epsilon)^2} \\
&+ C^{(7)}_x \frac{i}{x^2 - i\epsilon} + \cdots .
\end{split}
\label{eq:T_coosp}
\end{equation}
Here, $C^{(n)}_{x}$ are constants containing condensates with total mass dimension n and dimensionless numerical factors. 
In the above equation, the quark masses are taken as $m_u$ = $m_d$ = 0. 
This equation is only correct as long as we work at leading order in $\alpha_{s}$ for the Wilson coefficients, because higher
order corrections may involve additional logarithmic dependencies on $x^2$. 
The old-fashioned correlator can be derived from above equation using the Fourier transformation and the step function: 
\begin{equation}
\begin{split}
\Pi^{old}(q) = i \int d^4x e^{iqx} \theta (x_{0}) \Pi(x) \Big|_{\vec{q}=0}
\end{split}
\end{equation}
For example, in the dimension 5 term, one encounters the following equation: 
\begin{equation}
\begin{split}
&\displaystyle \int d^4x\theta(x_0) e^{iqx} \frac{i}{(x^2 - i\epsilon )^2} \Big|_{\vec{q}=0} = \\ 
&\displaystyle \int dx_0 \theta(x_0) e^{iq_0 x_0} \displaystyle \int d^3\vec{x} \frac{i}{(x_0^2 - \vec{x}^2 - i\epsilon)^2} 
e^{-i\vec{q}\cdot \vec{x}} \Big|_{\vec{q}=0}.
\end{split}
\end{equation}
First, we calculate the integrals over the spatial angles, leading to 
\begin{equation}
\begin{split}
&\frac{2\pi}{|\vec{q}|}
\displaystyle \int dx_0 \theta(x_0) e^{iq_0 x_0} \ \times \\
&\hspace{0.5cm}\displaystyle \int_{-\infty}^{\infty} dr 
\frac{r}{(r - x_0 + i\epsilon)^2 (r + x_0 - i\epsilon)^2} e^{i |\vec{q}|r} \Big|_{\vec{q}=0},
\end{split}
\label{eq:expl2}
\end{equation}
where we have used the definition $r \equiv |\vec{x}|$. Next comes the the integral over $r$, which can be treated 
in a standard way with the help of the Cauchy theorem. We thus obtain
\begin{equation}
\begin{split}
\pi^2 \displaystyle \int dx_0 \theta(x_0) \frac{1}{x_0 - i\epsilon} e^{ix_0(q_0 - |\vec{q}|)} \Big|_{\vec{q}=0} \\
=\frac{\pi}{2i} \displaystyle \int dk_0 \displaystyle \int dx_0 \frac{1}{k_0 - i \epsilon} \frac{1}{x_0 - i\epsilon} 
e^{ix_0(q_0 + k_0)}.
\end{split}
\label{eq:expl1}
\end{equation}
At this point, we have dropped $|\vec{q}|$, as there is no danger that the limit $|\vec{q}| \to 0$ leads 
to a divergence. 
Making use of Eq.(\ref{eq:F_step}) for the integral over $x_0$, we reach the final result: 
\begin{equation}
\pi^2 \displaystyle \int_{-q_0}^{\infty} dk_0 \frac{1}{k_0 - i\epsilon} = 
-\pi^2 \ln(-q_0 - i\epsilon) + \pi^2 \ln(\infty - i\epsilon).
\end{equation}
Here, we encounter a divergence in the second term, which, however, leads to no relevant contribution to the 
imaginary part of the correlator, which is the only quantity that is needed for the sum rules. We can therefore ignore 
it. Taking the imaginary part, we obtain the final result: 
\begin{equation}
\mathrm{Im} \Bigl [\int d^4x\theta(x_0) e^{iqx} \frac{i}{(x^2 - i\epsilon )^2} \Bigr ] \Bigg|_{\vec{q}=0} =  \pi^{3} \theta(q_{0}).
\end{equation}
The other terms of dimension 0, 3 and 4 can be calculated in a similar way. 
Hence, we here show only the results: 
\begin{equation}
\begin{split}
\mathrm{dim.\,0:\,}& \displaystyle \mathrm{Im} \Bigl [\int d^4x\theta(x_0) e^{iqx} \frac{\slash{x}}{(x^2 - i\epsilon )^5} \Bigr ] \Bigg|_{\vec{q}=0}\\
&= \frac{\pi^3}{2^9 3}\gamma_0 q_0^5 \theta (q_0), \\
\mathrm{dim.\,3:\,}& \displaystyle \mathrm{Im} \Bigl [ \int d^4x\theta(x_0) e^{iqx} \frac{i}{(x^2 - i\epsilon )^3} \Bigr ] \Bigg|_{\vec{q}=0}\\
&= -\frac{\pi^3}{8} q_0^2 \theta(q_0), \\ 
\mathrm{dim.\,4:\,}& \displaystyle \mathrm{Im} \Bigl [ \int d^4x\theta(x_0) e^{iqx} \frac{\slash{x}}{(x^2 - i\epsilon )^3} \Bigr ] \Bigg|_{\vec{q}=0}\\
&= \frac{\pi^3}{4}\gamma_0 q_0 \theta(q_0 ).
\end{split}
\label{eq:tekito1}
\end{equation}
We here, as above, have taken the limit $|\vec{q}| \to 0$. 
Note, however that in all the above calculations, this limit can be taken only after 
the integral over $r$ has been carried out, as otherwise the factor $1/|\vec{q}|$ appearing in 
Eq.(\ref{eq:expl2}) can not be properly treated.

For the other terms of dimension 6, 7 and 8, it is more convenient to work in momentum space 
where the time ordered correlator can be given as 
\begin{equation}
\begin{split}
\Pi(q) =& 
\big[ C^{(0)}_q q^4  \ln(-q^2 - i\epsilon) + C^{(4)}_q \ln(-q^2 - i\epsilon) \\
&+ C^{(6)}_q \frac{1}{q^2 + i\epsilon} 
+ C^{(8)}_q \frac{1}{(q^2 + i\epsilon)^2} + \dots \big]\slash{q} \\
&+ C^{(3)}_q q^2 \ln(-q^2 - i\epsilon) + C^{(5)}_q \ln(-q^2 - i\epsilon) \\
&+ C^{(7)}_q \frac{1}{q^2+ i\epsilon}+ \dots.
\end{split}
\label{eq:T_mosp}
\end{equation} 
Here, as in Eq.(\ref{eq:T_coosp}), 
$C^{(n)}_q$ contains only condensates and numerical factors. 
Furthermore, we here have neglected all polynomials in $q^2$, as they are not relevant for the present analysis. 
The old-fashioned correlator can now be derived from the momentum expression using the Fourier transformation and the step function as follows: 
\begin{equation}
\begin{split}
\Pi^{old}(q) = i \int d^4x e^{iqx} \theta (x_{0}) \int \frac{d^4p}{(2\pi)^4} e^{-ipx}\Pi(p) \Big|_{\vec{q}=0}.
\end{split}
\end{equation}
For the dimension 6 term, this form is the same as the phenomenological side of Eq.(\ref{eq:forward_phys1}). 
Thus, we obtain the final result immediately: 
 \begin{equation}
 \begin{split}
\displaystyle \mathrm{Im}\Bigl [\int d^4x\theta(x_0) e^{iqx} \displaystyle \int \frac{d^4p}{(2\pi)^4} e^{-ipx} \frac{\slash{p}}{p^2 + i\epsilon} \Bigr ] \Bigg|_{\vec{q}=0}\\
= -\gamma_0 \frac{\pi}{2}\delta(q_{0}). 
\end{split}
\label{eq:expl6}
\end{equation}
The other terms containing condensates with dimension 7 and 8 are transformed into the corresponding terms of the old-fashioned correlator through the same procedure. 
The only difference is that due to the larger powers in the denominator, the calculations when using the 
Cauchy theorem become more complicated. 
We here show only the corresponding results: 
\begin{equation}
\begin{split}
\mathrm{dim.\,7:\,}& \displaystyle \mathrm{Im}\Bigl [ \int d^4x\theta(x_0) e^{iqx} \displaystyle \int \frac{d^4p}{(2\pi)^4} e^{-ipx} \frac{1}{(p^2 + i\epsilon)^2} \Bigr ] \Bigg|_{\vec{q}=0}\\
&=\mathrm{Im}\Bigl [ \frac{1}{2} \frac{1}{|\vec{q}| - i\epsilon} \frac{1}{q_0 - |\vec{q}| + i\epsilon} \Bigr ] \Bigg|_{\vec{q}=0}, \\
\mathrm{dim.\,8:\,}& \displaystyle \mathrm{Im}\Bigl [ \int d^4x\theta(x_0) e^{iqx} \displaystyle \int \frac{d^4p}{(2\pi)^4} e^{-ipx} \frac{\slash{p}}{(p^2 + i\epsilon)^2} \Bigr ] \Bigg|_{\vec{q}=0}\\
&=\mathrm{Im}\Bigl [ \gamma_0 \frac{1}{4} \frac{1}{|\vec{q}| - i\epsilon} \frac{1}{(q_0 - |\vec{q}| + i\epsilon)^2} \Bigr ] \Bigg|_{\vec{q}=0}.
\end{split}
\end{equation}
As is clear from these expressions, we can at this point not take the limit $\vec{q} \rightarrow 0$
as it would lead to a divergence. This problem is only cured after the integral
over $q_{0}$ is carried out, as will be discussed in the next section. 

Next, we consider the $\alpha_{s}$ correction terms. 
For the nucleon sum rules, these corrections have been calculated for the Wilson coefficients of dimension 0, 3 and 6. 
They partly have the same analytic structure as Eqs.(\ref{eq:T_coosp}) and (\ref{eq:T_mosp}), which can be treated as explained above (we ignore the 
running of $\alpha_s$ here). However, for the dimension 0 and 6 terms, the $\alpha_s$ corrections involve contributions with additional logarithms, 
which complicate the calculation. 
In momentum space, these terms are expressed as 
\begin{equation}
\begin{split}
\mathrm{dim.\,0:\,\,}& \displaystyle  \slash{q} \ q^4  \bigl (\ln(-q^2 - i\epsilon) \bigr )^2, \\
\mathrm{dim.\,6:\,\,}& \displaystyle  \frac{\slash{q}}{q^2 + i\epsilon} \ln(-q^2 - i\epsilon). 
\label{eq:terms_alphas}
\end{split}
\end{equation}

For the term of dimension 0,  we use the equations given below \cite{Langwallner}: 
\begin{equation}
\begin{split}
&\int d^{4}x e^{iqx}\frac{\slash{x}}{(x^2)^5} =-\frac{\pi^{2}}{2^{9}\ 3}\slash{q} \ (q^2)^2 \ln(-q^2), \\
&\int d^{4}x e^{iqx}\frac{\slash{x}}{(x^2 - i\epsilon )^5} \ln(-x^{2}) =  \frac{\pi ^{2}}{2 ^{10}\ 3^{2}}\slash{q} \ (q^{2})^2 \\
& \hspace{1.5cm}\times \Bigl[ 3 \bigl (\ln (- q^2/4 ) \bigr )^2 + 2 (6\gamma -\frac{43}{4}) \ln ( -q^2/4) \Bigr ],
\end{split}
\end{equation}
where $\gamma$ is the Euler constant. 
Making use of these two equations, we first derive the $\alpha_{s}$ correction term in coordinate space. 
Then, we can calculate the old-fashioned correlator using Eq.(\ref{eq:tekito1}) and  
\begin{equation}
\begin{split}
& \mathrm{Im} \Bigl [ \int d^{4}x \theta (x_{0})e^{iqx}\frac{\slash{x}}{(x^2-i\varepsilon )^5} \ln \bigl(-(x^{2} - i\epsilon )\bigr) \Bigr ] \Bigg|_{\vec{q}=0} \\
&\hspace{1.8cm} =  - \gamma_{0}\frac{q_{0}^{5}\pi ^{3}}{2^{11}\ 3^{2}} \bigl(24\gamma -43 + 12\ln(q_{0}^{2}/4) \bigr) \theta (q_{0}).
\end{split}
\label{eq:tekito2}
\end{equation}
Note that for obtaining Eq.(\ref{eq:tekito2}) with the help of the Cauchy theorem, 
one must consider the complicated contour avoiding the branch cut of  the logarithmic function. 

As for the dimension 6 term, we can derive its old-fashioned correlator as follows: 
\begin{equation}
\begin{split} 
\displaystyle & \mathrm{Im}\Bigl [\int d^4x\theta(x_0) e^{iqx} \\
& \hspace{1.2cm} \times \displaystyle \int \frac{d^4p}{(2\pi)^4} e^{-ipx} \frac{\slash{p}}{p^2 + i\epsilon} \ln(-p^2 - i\epsilon) \Bigr ] \Bigg|_{\vec{q}=0} \\
 &= \gamma_0 \mathrm{Im}\Biggl [ \int^{\infty}_{-\infty} \frac{dq_{0}}{2\pi i} \frac{1}{q_{0} - i \epsilon} \\
 & \hspace{1.2cm} \times \frac{q_{0}+k_{0}}{\bigl( q_{0}+k_{0}  -|\vec{q}| + i \epsilon \bigr) \bigl(q_{0}+k_{0} +|\vec{q}| - i \epsilon  \bigr) }  \\
& \hspace{0.15cm} \times \Bigl ( \ln (|\vec{q}| -q_{0}-k_{0} - i \epsilon )  + \ln (|\vec{q}| +q_{0}+k_{0} - i \epsilon )  \Bigr )  \Biggr ] \Bigg|_{\vec{q}=0}. 
\end{split}
\label{eq:tekito3}
\end{equation}
In the first (second) term, there is a branch cut due to the logarithmic function below (above) the real axis. 
Therefore, we can choose to close the contour for each term such that it does not contain the cut and then simply use the Cauchy theorem. 
We thus obtain 
\begin{equation}
\begin{split}
\displaystyle &\mathrm{Im}\Bigl [\int d^4x\theta(x_0) e^{iqx} \displaystyle \int \frac{d^4p}{(2\pi)^4} e^{-ipx} \frac{\slash{p}}{p^2 + i\epsilon} \\ & \times \ln(-p^2 - i\epsilon) \Bigr ] \Bigg|_{\vec{q}=0} 
  = \gamma_0 \mathrm{Im}\Bigl [  \frac{1}{q_{0} + i\epsilon} \ln(-q_{0} -i\epsilon )\Bigr ] .
\label{eq:tekito4}
\end{split}
\end{equation}
In summary,  we show the relation between the time ordered correlator and the imaginary part of the old-fashioned 
correlator in the first two columns of Table \ref{tab:the correspondence relation}.

From the above procedure, we can obtain the old-fashioned correlator of the nucleon channel up to dimension 8 and first order $\alpha_{s}$ corrections 
from the time ordered correlator \cite{Ovchinnikov,KOhtani} as: 
\begin{equation}
\begin{split}
\Pi^{\mathrm{old}}(q_0) &= \gamma_0 \Pi_1^{\mathrm{old}}(q_0) + \Pi_2^{\mathrm{old}}(q_0), \\
\mathrm{Im}\biggl [\frac{\Pi ^{\mathrm{old}}_{1}(q_{0})}{\pi} \biggr ] \bigg |_{\vec q =0}
&=  (C_{0} + C_{0\alpha_{s}}) q_{0}^{5}  \theta (q_{0})   \\
&\quad +C_{4} \langle \frac{\alpha _s}{\pi }G^{2}\rangle q _{0}\ \theta (q_{0})  \\
                  & \quad  +C_{6} \langle \overline{q}q\rangle ^2  \delta (q_{0})  +C_{6\alpha_{s}} \langle \overline{q}q\rangle ^2 \\
                  & \quad +C_{8} \langle \overline{q}q\rangle \langle \overline{q}g\sigma \cdot Gq\rangle, \\
\mathrm{Im}\biggl [\frac{\Pi ^{\mathrm{old}}_{2}(q_{0})}{\pi}\biggr ] \bigg |_{\vec q =0}  
&= (C_{3} + C_{3\alpha_{s}}) \langle \overline{q}q\rangle q^{2}_{0} \ \theta (q_{0})  \\
 &\quad +C_{5} \langle \overline{q}g\sigma \cdot Gq\rangle \ \theta (q_{0}) \ \ \ \ \ \ \ \ \ \ \ \ \ \\
 &\quad + C_{7} \langle \overline{q}q\rangle \langle \frac{\alpha _s}{\pi }G^{2} \rangle.
\label{eq:Pi1_old}
\end{split}
\end{equation}
with 
\begin{equation}
\begin{split}
C_{0}&= \frac{5+2\beta +5\beta ^2 }{2^{11}\pi  ^4} \\
C_{0\alpha_{s}}&= \frac{5+2\beta +5\beta ^2 }{2^{11}\pi  ^4} \biggl[ \frac{71}{12} - \ln (\frac{q^2}{\mu ^{2}}) \biggr ] \frac{\alpha_s}{\pi} \\
C_{3}&= -\frac{1}{2^{6}\pi  ^2}\biggl [  7 -2\beta -5\beta ^2  \biggr ]\\
C_{3\alpha_{s}}&= -\frac{1}{2^{6}\pi  ^2}\biggl [  7 \ \frac{15}{14} -2\beta \ \frac{3}{2} -5\beta ^2 \ \frac{9}{10}  \biggr ] \frac{\alpha_s}{\pi} \\
C_{4}&= \frac{5+2\beta +5\beta ^2 }{2^{10}\pi ^2}\\
C_{5}&= \frac{3(1-\beta ^{2})}{2^{6}\pi ^2}\\
C_{6}&= \frac{1 }{2^{4}\ 3} \bigl [ 7 - 2\beta -5\beta ^2  \bigr ] \\
C_{6\alpha_s}&= \frac{1 }{2^{4}\ 3} \biggl [ 7 \ \frac{325}{126} + 2\beta \frac{224}{9} 
              + 5\beta ^2  \ \frac{511}{90}   \biggr ] \frac{\alpha_s}{\pi} \delta (q_{0}) \\
 & \quad - \frac{1 }{2^{3}\ 3 \pi}\biggl[ 7 \  \frac{47}{21} - 2\beta \ \frac{5}{3} 
                  -5\beta ^2 \ \frac{61}{15} \biggr ] \frac{\alpha_s}{\pi} \\
& \quad \times \mathrm{Im}\biggl [ \frac{1}{q_{0} + i\epsilon } \ln \Bigl (-\frac{q_{0}+ i\epsilon}{\mu} \Bigr ) \biggr ]\\
C_{7}&= \frac{1}{\pi} \cdot \frac{19 + 10\beta -29\beta ^2 }{2^{8}\ 3^{2}} \times \\ 
       & \quad \mathrm{Im} \biggl [ \frac{1}{ |\vec q| -i\epsilon } \cdot \frac{1}{(q_{0}-|\vec q|+i\epsilon )} \biggr ] \bigg |_{\vec q =0} \\
C_{8}& = -\frac{1}{\pi} \cdot \frac{13 - 2\beta -11\beta ^2 }{2^{7}\ 3} \times \\ 
      & \quad \mathrm{Im}\bigg [ \frac{1}{| \vec q|-i\epsilon}
\frac{1}{(q_{0}-|\vec q|+i\epsilon )^{2}} \bigg ] \bigg |_{\vec q =0},
\end{split}
\end{equation}
where $\mu$ is a renormalization constant.
In the equations above, we have omitted the terms with the gluonic condensates of dimension 6 and 8 ($\langle G^3 \rangle$, $\langle G^4 \rangle$). 
As we will explain in the next subsection, 
these terms will vanish at leading order in $\alpha_s$ for the kernel used in this paper and we thus do not consider them in the present analysis. 
For a more general investigation of the nucleonic sum rules, they, however, should in principle be taken into account. 

\subsection{Construction of the sum rule}
\label{app:2}
In this subsection, we show how the final form of the sum rule is constructed, starting from Eq.(\ref{eq:finalpar}) and the 
results obtained above. Here, we assume a general form of the kernel, which we take to be either an even (odd) function of $q_0$ 
and denote it as $W_{e}(q_0^2)$ ($q_0W_{e}(q_0^2)$), where $W_{e}(q_0^2)$ is analytic in the upper half of the imaginary plane and real on the real axis. 
Furthermore, we will prove that 
the potential divergences of the higher order terms of dimension 7 and 8 vanish through the integral of Eq.(\ref{eq:finalpar}). 

Firstly, we consider the terms up to dimension 5. For these, the imaginary parts of the old-fashioned correlator is 
proportional to a step function and one therefore just has to substitute our findings of the last subsection into Eq.(\ref{eq:finalpar}) and evaluate the integral from 
0 to $\infty$. This leads to the results presented in Eq.(\ref{eq:OPEfres}). 

Next, let us discuss the more complicated terms 
of dimension 6, 7 and 8. 
As a first step, we show that, if one takes the limit $|\vec{q}| \to 0$, 
the imaginary parts of the leading terms of dimensions 6,8,... (7,9,...) are even (odd) functions of $q_0$. 
By following the procedure of the last subsection, 
it is noticed that all terms appearing at dimensions 6,8,... can generally be written down as 
\begin{equation}
\begin{split}
F_1(q_0) &\equiv  \lim_{\vec{q} \to 0} \frac{1}{2\pi i} \displaystyle \int _{-\infty} ^{\infty}dk_0 \frac{1}{k_0 - i\epsilon} \frac{k_0 + q_0}{[(k_0 + q_0)^2 -\vec{q} ^2+ i\epsilon]^n}  \\
&=\lim_{\vec{q} \to 0} \frac{1}{2\pi i} \displaystyle \int _{-\infty} ^{\infty} dk_0 \frac{1}{k_0 - i\epsilon} \times \\
& \hspace{1.0cm} \frac{k_0 + q_0}{(k_0 + q_0 - |\vec{q}| + i\epsilon)^n (k_0 + q_0 + |\vec{q}|- i\epsilon)^n} .\\
&\hspace{5.5cm} (n=1,2,\dots)
\label{eq:F1}
\end{split}
\end{equation}
Here, we are ignoring any proportional real constant, including $\gamma_0$. Similarly, for dimensions 7,9,..., 
we get 
\begin{equation}
\begin{split}
F_2(q_0) & \equiv \lim_{\vec{q} \to 0} \frac{1}{2\pi i} \displaystyle \int _{-\infty} ^{\infty} dk_0 \frac{1}{k_0 - i\epsilon} \frac{1}{[(k_0 + q_0)^2 -\vec{q} ^2 + i\epsilon]^n}  \\
&=\lim_{\vec{q} \to 0} \frac{1}{2\pi i} \displaystyle \int _{-\infty} ^{\infty} dk_0 \frac{1}{k_0 - i\epsilon} \times \\
&\hspace{1.0cm} \frac{1}{(k_0 + q_0 -|\vec{q}|+ i\epsilon)^n (k_0 + q_0 + |\vec{q}|- i\epsilon)^n}. \\
&\hspace{5.5cm} (n=1,2,\dots)
\label{eq:F2}
\end{split}
\end{equation} 
Next, we take the imaginary parts and, after some simple manipulations, get for $F_1(q_0)$ 
\begin{equation}
\begin{split}
\mathrm{Im} F_1(q_0) &= \frac{1}{2i} \Big[F_1(q_0) - \overline{F_1(q_0)}\Big] \\
&= -\frac{1}{4\pi} \lim_{\vec{q} \to 0} \displaystyle \int _{-\infty} ^{\infty}dk_0 \frac{1}{k_0 - i\epsilon} \times \\  
& \ \ \Big[ \frac{k_0 + q_0}{(k_0 + q_0 -|\vec{q}| + i\epsilon)^n (k_0 + q_0 + |\vec{q}|- i\epsilon)^n}\\ 
&\ \ \ -\frac{-k_0 + q_0}{(k_0 - q_0 + |\vec{q}|+ i\epsilon)^n (k_0 - q_0 -|\vec{q}|- i\epsilon)^n} 
\Big],
\end{split}
\end{equation}
while the result for $F_2(q_0)$ is 
\begin{equation}
\begin{split}
\mathrm{Im} F_2(q_0) &= \frac{1}{2i} \Big[F_2(q_0) - \overline{F_2(q_0)}\Big] \\
&= -\frac{1}{4\pi} \lim_{\vec{q} \to 0} \displaystyle \int _{-\infty} ^{\infty} dk_0 \frac{1}{k_0 - i\epsilon} \times \\ 
& \ \ \Big[ \frac{1}{(k_0 + q_0 -|\vec{q}|+ i\epsilon)^n (k_0 + q_0 +|\vec{q}| - i\epsilon)^n}\\
& \ \ \  -\frac{1}{(k_0 - q_0 +|\vec{q}| + i\epsilon)^n (k_0 - q_0 -|\vec{q}| - i\epsilon)^n} 
\Big].
\end{split}
\end{equation}
In the above Equations, we keep the $\vec{q}^2$ to clarify the sign of the imaginary part of the pole positions. 
Having the above equations at hand, it is now a trivial matter to show that 
\begin{equation}
\mathrm{Im} F_1(-q_0) = \mathrm{Im} F_1(q_0), 
\end{equation}
and 
\begin{equation}
\mathrm{Im} F_2(-q_0) = -\mathrm{Im} F_2(q_0), 
\end{equation}
which proofs that 
the imaginary parts of the terms of dimension 6,8,... (7,9,...) in the 
OPE of the old-fashioned correlator are even (odd) functions of $q_0$ in the limit $|\vec{q}| \to 0$.

Therefore, we can immediately evaluate half of the contributions of the these terms in Eq.(\ref{eq:finalpar}). 
Namely, we find that the leading terms of dimension 7,9,... vanish if  the kernel is an even function while the terms of dimension 6,8,10,...
vanish if it is an odd function. 

Among the remaining terms, the leading dimension 6 term with an even kernel can be obtained trivially, as the 
the imaginary part of the old-fashioned correlator is proportional to a delta function at the origin. 
To calculate the non-vanishing part of dimension 7, we consider the case where $W(q_0)$ is an odd function: 
\begin{equation}
\begin{split}
\mathrm{Im}\Bigl [ \int _{-\infty} ^{\infty} dq_0 \ \frac{1}{2} \frac{1}{|\vec{q}| - i\epsilon} \frac{1}{q_0 - |\vec{q}| + i\epsilon}  \cdot \ q_0 W_{e}(q_0^2) \Bigr ] \Big|_{\vec{q}=0} \\ 
= -\frac{\pi}{2}  W_{e}(0). 
\label{eq:expl5}
\end{split}
\end{equation}
It can be understood here that the seemingly divergent part $1/\vec{q}$ disappears after the final integration over $q_0$. 
For dimension 8,  we consider the case where $W_{e}(q_0)$ is an even function:
\begin{equation}
\begin{split}
&\mathrm{Im}\Bigl [ \int _{-\infty} ^{\infty} dq_0  \frac{1}{4} \frac{1}{|\vec{q}| - i\epsilon} \frac{1}{(q_0 - |\vec{q}| + i\epsilon)^2} W_{e}(q_0^2) \Bigr ] \Big|_{\vec{q}=0}  \\
&=\mathrm{Im}\Bigl [ \int _{-\infty} ^{\infty} dq_0  \frac{1}{4} \frac{1}{|\vec{q}| - i\epsilon} \frac{1}{(q_0 - |\vec{q}| + i\epsilon)} \frac{d}{dq_0}W_{e}(q_0^2) \Bigr ] \Big|_{\vec{q}=0} \\
&= -\frac{\pi}{2}  \frac{d}{d(q_0^2)}W_{e}(0). 
\label{eq:expl8}
\end{split}
\end{equation}
After integrating by parts, which leads to the second line in the above equation, we see that, as for the dimension 7 case, 
the diverging term vanishes. 
We have explicitly checked that for the dimension 9 term, all potential divergences vanish in a similar fashion. Thus, 
the same procedure could presumably be continued to even higher orders, but this is not of much practical use 
as the higher order terms have large numerical uncertainties from the corresponding condensates at present.
 
Finally, let us discuss the part of the $\alpha_{s}$ correction term of dimension 6, which contains a 
logarithm. Due to this logarithm, the contributions of this term using both $W_e(q_0^2)$ and $q_{0}W_e(q_0^2)$ are finite 
since this term is neither an even nor an odd function of $q_0$. 
For $W_e(q_0^2)$, the contribution can be calculated with the help of the Cauchy theorem by considering the contour avoiding the pole. 
The result can be obtained as follows:
\begin{equation}
\begin{split}
&\mathrm{Im}\Bigl [ \int _{-\infty} ^{\infty} dq_0  \frac{1}{q_{0} + i\epsilon } \ln \bigl (-q_{0}- i\epsilon \bigr ) W_{e}(q_0^2) \Bigr ]  \\
& = -\pi \ln(\epsilon) W (\epsilon^2) - \pi \int_{\epsilon}^{\infty} dq_0 \frac{1}{q_0} W_{e}(q_0^2) \\
& = 2 \pi \int _{0}^{\infty}dq_0 q_{0} \ln(q_{0}) \frac{dW_{e}(q_0^2)}{dq_0^2}.
\label{eq:expl6a}
\end{split}
\end{equation}
Note that in the second line, $\epsilon$ stands for the infinitesimal radius of the path around the origin of the imaginary plane and that the 
diverging parts of the two terms appearing there exactly cancel, leaving only the finite result of the third line. 

For $q_0W_{e}(q_0^2)$, the contribution can be immediately obtained as: 
\begin{equation}
\begin{split}
&\mathrm{Im}\Bigl [ \int _{-\infty} ^{\infty} dq_0  \frac{1}{q_{0} + i\epsilon } \ln \bigl (-q_{0}- i\epsilon \bigr ) q_0 W_{e}(q_0^2) \Bigr ]  \\
& = -\pi \int _{-\infty} ^{\infty} dq_0 \ \theta (q_{0})  W_{e}(q_0^2) \\ 
& = -\pi \int _{0} ^{\infty} dq_0 W_{e}(q_0^2). 
\end{split}
\end{equation} 
All the results of this subsection, using both even and odd kernels, are summarized in the third and fourth column of 
Table \ref{tab:the correspondence relation}. 
\\
\\
\begin{table*}[tbp]
\scalebox{0.99}{
\rotatebox{90}{\begin{minipage}{\textheight}
\begin{center}
\renewcommand{\arraystretch}{2.5}
\begin{tabular}{|p{4cm}|p{6cm}|p{6cm}|p{6cm}|}
\hline 
\makebox[4cm][c]{ $\Pi (q)$} & \makebox[6cm][c]{ $\mathrm{ Im}\bigl[\Pi ^{old} (q_{0})\bigr]$} &\makebox[6cm][c]{\centering $G(x)$ using $W _{e}(q_0 ^2)$} &
\makebox[6cm][c]{ $G(x)$ using $q_{0}W_{e} (q_0 ^2)$} \\ \hline
\makebox[4cm][c]{$\slash{q}\ (q^{2})^2 \ln(-q^2) $}&\makebox[6cm][c]{$-\gamma _{0}\pi q_{0}^{5} \theta (q_{0}) $}&\makebox[6cm][c]{ $-\gamma _{0} \int _{0}^{\infty}dq_0 q_{0}^{5} W_{e}(q_0^2) $} &\makebox[6cm][c]{ $-\gamma _{0}  \int _{0}^{\infty}dq_0 q_{0}^{6} W_{e}(q_0^2) $} \\ \hline
\makebox[4cm][c]{$\slash{q}\ (q^{2})^2 \bigl (\ln (- q^2/\mu ^2 ) \bigr )^2 $}&\makebox[6cm][c]{$-\gamma _{0}2\pi q_{0}^{5} \ln ( q_{0}^2/\mu ^2 ) \theta (q_{0}) $}&\makebox[6cm][c]{$-2\gamma _{0} \int _{0}^{\infty}dq_0 q_{0}^{5} \ln ( q_{0}^2/\mu ^2 ) W_{e}(q_0^2) $}
&\makebox[6cm][c]{$-2\gamma _{0} \int _{0}^{\infty}dq_0 q_{0}^{6} \ln ( q_{0}^2/\mu ^2 ) W_{e}(q_0^2) $} \\ \hline
\makebox[4cm][c]{$ q^{2} \ln(-q^2) $}&\makebox[6cm][c]{$-\pi q_{0}^{2} \theta (q_{0}) $}&\makebox[6cm][c]{$- \int _{0}^{\infty}dq_0  q_{0}^{2}  W_{e}(q_{0}^2) $}&\makebox[6cm][c]{$- \int _{0}^{\infty}dq_0  q_{0}^{3}  W_{e}(q_{0}^2) $} \\ \hline
\makebox[4cm][c]{$\slash{q} \ln(-q^2) $}&\makebox[6cm][c]{$-\gamma _{0}\pi q_{0} \theta (q_{0}) $}&\makebox[6cm][c]{$-\gamma _{0} \int _{0}^{\infty}dq_0 q_{0} W_{e}(q_{0}^2) $}&\makebox[6cm][c]{$-\gamma _{0} \int _{0}^{\infty}dq_0 q_{0}^{2} W_{e}(q_{0}^2) $} \\ \hline
\makebox[4cm][c]{$ \ln(-q^2) $}&\makebox[6cm][c]{$-\pi \theta (q_{0})$}&\makebox[6cm][c]{ $- \int _{0}^{\infty}dq_0 W_{e}(q_{0}^2)$}&\makebox[6cm][c]{ $- \int _{0}^{\infty}dq_0 q_{0}W_{e}(q_{0}^2)$} \\ \hline 
\makebox[4cm][c]{$\frac{q\hspace{-.35em}/}{q^2 + i\epsilon}$} &\makebox[6cm][c]{$-\ \gamma _{0}\frac{\pi}{2}\delta(q_{0})$}&\makebox[6cm][c]{$-\ \gamma _{0}\frac{1}{2}W_{e}(0)$}&\makebox[6cm][c]{0} \\ \hline 
\makebox[4cm][c]{$\frac{q\hspace{-.35em}/}{q^2 + i\epsilon} \ln(-q^2 - i\epsilon) $} &\makebox[6cm][c]{$\mathrm{Im}\Bigl [  \frac{\gamma _{0}}{q_{0} + i\epsilon} \ln(-q_{0} -i\epsilon )\Bigr ]$}&\makebox[6cm][c]{$2 \gamma _{0}  \int _{0}^{\infty}dq_0 q_{0} \ln(q_{0}) W'_{e}(q_0^2)$}
&\makebox[6cm][c]{$-\gamma_0 \int _{0}^{\infty}dq_0 W_{e}(q_{0}^2)$} \\[2pt]  \hline 
\makebox[4cm][c]{$\frac{1}{q^2 + i\epsilon}$}&\makebox[6cm][c]{$\mathrm{Im}\Bigl [ \frac{1}{2} \frac{1}{|\vec{q}| - i\epsilon} \frac{1}{q_0 - |\vec{q}| + i\epsilon} \Bigr ] \Big|_{\vec{q}=0}$}&\makebox[6cm][c]{0}&\makebox[6cm][c]{$-\frac{1}{2} W_{e}(0)$} \\[2pt]  \hline 
\makebox[4cm][c]{$\frac{q\hspace{-.35em}/}{(q^2 + i\epsilon)^2}$} &\makebox[6cm][c]{$\mathrm{Im}\Bigl [ \frac{1}{4} \frac{\gamma _{0}}{|\vec{q}| - i\epsilon} \frac{1}{(q_0 - |\vec{q}| + i\epsilon)^2} \Bigr ] \Big|_{\vec{q}=0}$}&\makebox[6cm][c]{$-\gamma _{0} \frac{1}{2} W'_{e}(0)$}&\makebox[6cm][c]{0} \\[2pt] 
\hline
\end{tabular}
\caption{The relation between each term of the time ordered correlator $\Pi(q)$, the imaginary part of the old-fashioned correlator $\Pi ^{old}(q_{0})$ and the integration values $G(x)$ using 
$W_e(q_0^2)$ and $q_0W_e(q_0^2)$. Here, x stands for the parameters appearing in $W(q_{0})$, such as Borel mass $M$ or $s$ and $\tau$. 
The first column shows the terms appearing in the time ordered correlator and 
the second column gives the imaginary parts of the corresponding terms of 
the old-fashioned correlator with $\vec{q} = 0$. 
The third and fourth columns give the integration values using $W_{e}(q_0^2)$ and $q_0W_{e}(q_0^2)$ respectively. 
Here, $W'_{e}(q_0^2)$ stands for $\frac{dW_{e}(q_0^2)}{d(q_0^2)}$.} 
\label{tab:the correspondence relation}
\end{center}
\end{minipage}
}
}
\end{table*}

\end{document}